\documentclass[twocolumn,pra,showpacs,floatfix,amsmath,amssymb,amsfonts,aps]{revtex4-1}
\usepackage{graphicx} 
\usepackage{subfigure}
\usepackage{dcolumn}
\usepackage{tabularx}
\usepackage[usenames]{color}
\newcommand{\la}{\left\langle}
\newcommand{\ra}{\right\rangle}
\newcommand{\eg}{{e.g.}\ }
\newcommand{\ie}{{i.e.}\ }
\newcommand{\etal}{\emph{et al.}\ }

\newcommand{\beq}{\begin{equation}}
\newcommand{\eeq}{\end{equation}}
\newcommand{\beqa}{\begin{eqnarray}}
\newcommand{\eeqa}{\end{eqnarray}}
\newcommand{\beqn}{\begin{equation*}}
\newcommand{\eeqn}{\end{equation*}}
\newcommand{\beqan}{\begin{eqnarray*}}
\newcommand{\eeqan}{\end{eqnarray*}}

\newcommand{\ar}{\begin{array}}
\newcommand{\ear}{\end{array}}
\newcommand{\bc}{\begin{color}}
\newcommand{\ec}{\end{color}}
\newcommand{\bit}{\begin{itemize}}
\newcommand{\eit}{\end{itemize}}

\begin{document}

\title{The Multi-Configurational Hartree-Fock close-coupling ansatz:\\
application to Argon photoionization  cross section and delays}

\author{T. Carette$^1$, J. M. Dahlstr\"om$^1$, L. Argenti$^2$ and E. Lindroth$^1$}\email{Eva.Lindroth@fysik.su.se}
\affiliation{$^1$ Department of Physics, Stockholm University, AlbaNova University Centre, SE-106 91 Stockholm, Sweden}
\affiliation{$^2$ Departamento de Qu\'imica, M\'odulo 13, Universidad Aut\'onoma de Madrid, 28049 Madrid, Spain}

\date{\today}

\begin{abstract}
We present a robust, \emph{ab initio} method for addressing atom-light interactions and apply it to photoionization of argon. We use a close-coupling ansatz constructed on a multi-configurational Hartree-Fock description of localized states and B-spline expansions of the electron radial wave functions. In this implementation, the general many-electron problem can be tackled thanks to the use of the ATSP2K libraries [CPC {\bf 176} (2007) 559]. In the present contribution, we combine this method with exterior complex scaling, thereby allowing for the computation of the complex partial amplitudes that encode the whole dynamics of the photoionization process.
The method is validated on the $3s3p^6np$ series of resonances converging to the $3s$ extraction. Then, it is used for computing the energy dependent differential atomic delay between $3p$ and $3s$ photoemission, and agreement is found with the measurements of Gu\'enot \etal~[PRA {\bf 85} (2012) 053424]. The effect of the presence of resonances in the one-photon spectrum on photoionization delay measurements is studied.
\end{abstract}

\maketitle

%
\section{Introduction}
Since its discovery, the photoelectric effect has
occupied a forefront position among the processes triggered by the
interaction between matter and radiation.  This is due to the relevance
of photoionization for a range of naturally occurring phenomena, as for
example those determining the opacity of astrophysical
objects~\cite{badnell:2005}, or leading to radiation damage of
biological systems~\cite{Boudaiffa03032000}, as well as for many
technological applications.  A quantitative description of
photoionization is required to understand and control these processes.
Research in this direction has a long
history~\cite{fano:1968,starace:1980} and generated a vast literature.
The most recent development in this area includes the advent of several
new light sources providing shorter pulses, higher intensities or
shorter wave lengths~\cite{Sansone2011,Young:nature:10}. During the last years, such sources
have been utilized to gain knowledge also of the temporal aspects of
photoionization~\cite{Schultze2010a,klunder:2011}.

A reliable description of the photoionization event requires a good
representation of the atomic or molecular system during all the stages
of its interaction with light. This includes the structure of the initial bound state, 
as well as of the final {\em parent} or {\em target} states, but also the coupling between the 
released photoelectron and the relaxing parent ion. 
 In this work, we present a tool that permits to treat
many-body effects in bound and continuum states of many-electron
atoms, irrespective of how the photoionization process is to be 
treated. Depending on the intensity, pulse length and monochromaticity  of the light 
the appropriate method  may be found everywhere  on the range from the standard one-photon time-independent formalism, 
to the explicit solution of the time-dependent Schr\"odinger equation.
Since the latter approach quickly grows computationally heavy for all
but the smallest systems, it is usually
necessary to stick to a simplified treatment of correlation for larger systems.
The procedure we outline here allows for  a  systematic refinement of many-body effects and 
the unified  approach for time-independent and time-dependent calculations 
permits  further tests and   verifications of the amount of correlation  in the latter by comparison with the effect on the former.

In the present study we 
 validate the approach and verify that the underlying
approximations are justified by comparing photoabsorption cross
sections for the argon atom, computed in the weak field limit, with
existing experimental and theoretical data.  Furthermore, we show that
the delay in the photoelectron ejection observed experimentally~\cite{klunder:2011,Gueetal:12a} can be largely explained by the single photoionization scattering phases when corrected for the phase shift introduced 
in the experiment by the probe photon~\cite{marcus:tutorial:2012}. 

For the description of the wave
functions in the continuum two different approaches dominate the literature. 
One possibility is to match a numerical solution, computed
inside a inner spherical region, with the appropriate asymptotic solution in
the outer region, whose analytical form is known. This is the basis
of most scattering methods like, e.g., the
R-matrix method, originally adapted for the
photoionization problem by Burke and Taylor~\cite{BurTay:75a} and
recently presented also in a time-dependent
version~\cite{timedependentRmatrix:2012},  as well as the K-matrix
method~\cite{moccia:1991,argenti_moccia:2010}.
An alternative solution is to use complex scaling, where radial
coordinates are rotated in the complex plane.  In this case, the
photoabsorption cross section can be directly computed as the imaginary
part of the frequency-dependent
polarizability~\cite{rescigno:75:cr,lindroth:95:liminus,chung:li:1997}.
Complex scaling methods present some advantages as well as some
disadvantages with respect to scattering methods. On the one hand, with
complex scaling, resonance states have a more transparent representation
since they appear directly as eigenstates of the complex-scaled
Hamiltonian with a negative imaginary part of the energy. On the other
hand, the complex-scaled wave functions differ from the generalized
eigenstates of the unscaled Hamiltonian. Therefore, properties that can
only be extracted from the ionized part of the wave function are not
easily obtainable.  This problem is circumvented using exterior complex
scaling~\cite{simon:79}, where the radial coordinate is scaled only
outside a sufficiently large region. In this method, the wave function
in the inner unscaled region can be directly analysed. Exterior complex
scaling has been extensively applied to electron and photon scattering
problems~\cite{mccurdy:97,mccurdy:97:longrange,horner:twophoton:2007,rescigno:h2:07},
see~\cite{McCurdy2004b} for a review.

The description of the target and parent-ion states is an atomic structure problem than can be treated with
the tools targeting correlation in bound states, e.g. many-body
perturbation theory, coupled-cluster,
configuration interaction, etc., see e.g.~\cite{mbpt,Froetal:97a,johnsonbook}. 
Most of these approaches can
subsequently be combined with different representations of the
continuum.  Configuration interaction, for example, can be used to
either represent the atomic system in the inner R-matrix region or be
employed together with complex scaling.  A method that is particularly
convenient when aiming at problems involving single atomic continua, but
non-trivial configurations, is the Multi-Configuration Hartree-Fock
(MCHF) method~\cite{Fro:77a,Froetal:97a}. It is known to yield
accurate correlation models for general bound states~\cite{FroJon:94a}
while still limiting the number of configurations required for convergence.
 MCHF has often been
used in combination with the R-matrix method, see e.g.
Ref.~\cite{Beretal:96a}, but also together with other approaches to describe the continuum (see \eg Ref.~\cite{manson:97}).
	  Here, we employ the close-coupling
	ansatz, originally introduced by Massey and Mohr~\cite{massey:32,massey:33}, for the atomic wave function. 
	 The localized part of the wave function is expressed
	in terms of MCHF configurations; our  implementation is based
	on the MCHF atomic-structure package ATSP2K~\cite{Froetal:07a}, while the
	radial component that corresponds to the
	photoelectron is expanded on an extensive B-spline basis. We
	use exterior complex scaling for an explicit description of resonances
	and a good representation of the continuum.

\begin{figure}
\caption{Diagram of the neutral argon levels and ionization thresholds accessible through one-photon transitions.
Data are taken from the NIST atomic database~\cite{nist_13_01_08}.
The red level is the Ar ground state, and the blue states are singly excited bound and autoionizing states. Black lines are ionization thresholds that are included in our analysis. Grey lines are other ionization thresholds which are neglected in the present paper (cf. Appendix~\ref{appendix}).
\label{fig:levels}}
\includegraphics[width=\columnwidth]{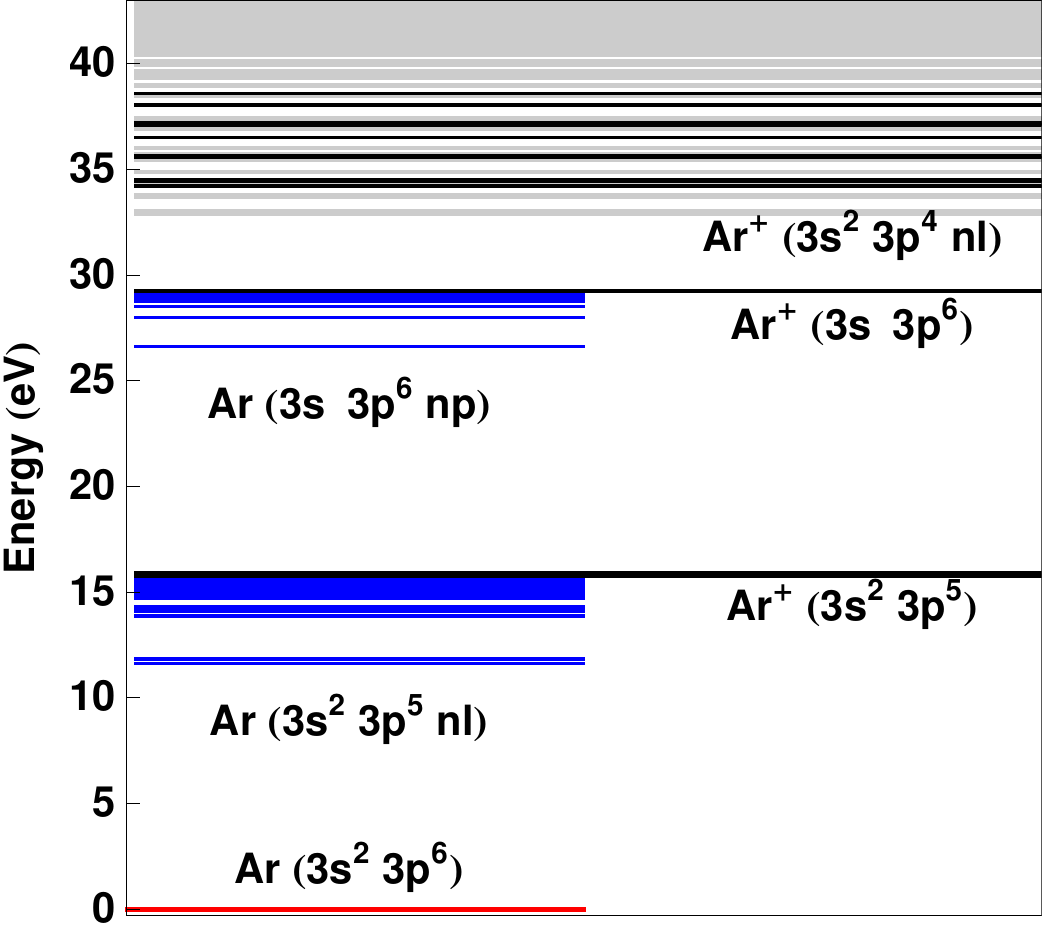}
\end{figure}

This paper is organized as follow. In Sec.~\ref{sec2}
we outline the present implementation of the close-coupling ansatz, and in 
Sec.~\ref{sec3} the calculation of photoionization parameters is discussed.
To validate the method,  we
compare our results for the photoionization of the argon atom with other
theoretical and experimental data in Sec.~\ref{sec:res}. Argon provides a good benchmark because it
is comparatively light, so a non-relativistic description is
appropriate, and its resonance profiles and angular distribution, which
encode the full information about the complex partial
amplitudes~\cite{KabSaz:76a}, have been the target of many experimental
and theoretical
studies~\cite{BurTay:75a,AmuKhe:82a,Soretal:94a,Madetal:69a,Wuetal:95a,Beretal:96a,Tay:77a,Codetal:80a,Sveetal:87a,Beretal:96a}.
Fig.~\ref{fig:levels} is a diagram of the levels of neutral argon accessible through one-photon transition, showed on an energy scale corresponding to photoionization from the ground state. Above the, $3s3p^6$ threshold, we only mark the levels of Ar$^+$ in black and grey.
In Sec.~\ref{sec:res}, we focus our attention on the photon energy range between 26~eV and
30~eV which comprises the $3s 3p^6 np$ autoionizing series, which entails
strong configuration interaction.
In Sec.~\ref{sec:delay} we use our
approach to estimate the one-photon Wigner-like delay in the experimentally relevant region from 32~eV to 42~eV. The importance of the  presence of many resonances affecting the atomic delay is demonstrated.
Finally, in
Sec.~\ref{sec:conclusions} we draw our conclusions.

\section{The close-coupling ansatz}\label{sec2}
\subsection{Exterior complex scaling}

In our implementation of the close-coupling ansatz, the states in the
continuum are evaluated on an exterior complex scaled radial coordinate

\beq
 \label{exterior} r \rightarrow r^{(\theta)} \equiv \left\{ \ar{ll}
r & \text{if } r < R_0 \\
R_0 + e^{i\theta}(r-R_0) & \text{if } r\geq
R_0 \, . \ear \right.
\eeq
 As mentioned earlier, this approach provides a
good description of the photo-absorption cross section in general, and
of transiently-bound states in particular.  In the present context, ECS
has two further advantages.  First, as detailed in
Sec.~\ref{close-coupling ansatz}, we employ existing MCHF
codes~\cite{Froetal:07a} to describe the bound states. If $R_0$ is large
enough, the output of these programs can be directly used in conjunction
with additional complex-scaled functions without any modification. In
fact, this was the reason why Simon~\cite{simon:79}, who wanted to treat
molecules without having to scale the molecular potential, introduced
ECS to start with.  Second, with ECS we are able to analyze the ionized
part of the wave function to determine partial complex amplitudes. This
will be discussed more in detail in Sec.~\ref{sec3}.

\subsection{Multi-Configuration Hartree-Fock } \label{close-coupling ansatz}
In the MCHF method, the atomic wave
function is expanded on an orthonormal set of symmetry adapted linear
combinations of Slater determinants~\cite{Fro:77a,Froetal:97a}, $ \{
	\Phi (\gamma_i LS) \} $, known as Configuration State Functions (CSF),
\begin{equation}
 \Psi = \sum_j c_j \Phi (\gamma_j LS).
\end{equation}
The radial functions $\{ P_{nl}(r) \}$, which define the occupied and
	correlation spin-orbitals, and the  $\{c_j \}$ mixing coefficients are
	then optimized to minimize the energy functional 
\beq
	\hspace{-1.5cm} E[\{ P_{nl}(r) \},\{c_j\}] = \la \Psi \left| H \right| \Psi\ra
\eeq
where $H$ is the non-relativistic atomic Hamiltonian with infinite mass nucleus, 
under the orthogonality constraint for the radial orbitals $\{ P_{nl}(r) \}$:
\beq
\label{orth} \int\!\! P_{nl}(r)P_{n'l}(r) dr = \delta_{nn'}\ .  
\eeq
 We use the ATSP2K package~\cite{Froetal:07a} to
	solve the MCHF problem.

The close-coupling expansion is defined by a set of $N$-electron target
states ($\Psi_a$) coupled to an extensive set of one-particle states
with defined orbital angular momentum $l$. This group of $(N+1)$-particle
states is further complemented by a set of localized states
($\chi_\xi$).  The wave function of the atomic state before or after
photo-absorption can  be expressed as 
\beq
\label{ciansatz} \Psi =
\sum_\xi c_\xi \chi_\xi + \sum_{\alpha,n} c_{\alpha n}\
\mathcal{A}_{LS}\{\Psi_a P_{nl}(r_{N+1})\} 
\eeq
 where $\alpha$ is the
channel index, indicating target state $a$ and photoelectron angular
momentum $l$, and $\mathcal{A}_{LS}$ yields a symmetry adapted,
antisymmetric $(N+1)$ - electron wave function.  The $N$-electron CSF
are built from the orbitals obtained in the MCHF calculation, and we
find the target states ($\Psi_a$) as the linear combinations that
diagonalize the total $N$-electron Hamiltonian.  We further use these
orbitals to set up the $(N+1)$-electron CSF that form the $\chi_{\xi}$.  All
intermediate angular couplings compatible with the total symmetry under
examination are included in the close-coupling expansion which, in this
sense, is therefore complete.  As in MCHF-based R-Matrix
approaches~\cite{Beretal:96a}, orbitals that are occupied in the
target states, and which derive from a MCHF optimization, are kept
frozen throughout the calculations.  A discretized  description of the
continuum orbitals $P_{nl}(r)$ describing the photoelectron is assumed.  The CI coefficients, $c_\xi$ and $c_{\alpha n}$
in Eq.~(\ref{ciansatz}) give the close-coupling ansatz a great flexibility.
Even if occupied orbitals are optimized on the parent states, the second
term of Eq.~(\ref{ciansatz}) is able to account for the orbital
adjustment between the $N$- and $N+1$-electron system.
One drawback of this approach is that highly excited configurations or configurations involving correlation orbitals often appear as discrete states that are embedded in the continua, slightly above the physical states described by the model. These states appear as artifacts of the calculation  in the photoionization cross-section, so-called pseudo-resonances.

\subsection{Description of the photoelectron} \label{bsplines} The
photoelectron radial function basis $\{P_{nl}\}$, cf. Eq.~(\ref{ciansatz}), is expanded in
B-splines~\cite{deboor}
 
\beq
 P_{nl} (r^{(\theta)}) =  \sum_i c_i B_i^k
(r^{(\theta)})\, .  \label{pelB} 
\eeq
 The B-splines of order $k$ are
defined as  piecewise polynomials of order $k-1$ between pre-defined isolated
points, called knots. In this work, we use k=7. With exterior complex scaling, it
is convenient to let the knot sequence follow the scaled $r$ in
Eq.~(\ref{exterior}), \ie $r^{(\theta)}$~\cite{ecs:Bslines:2004}.
B-splines can be used to form effectively complete basis sets for the
description of localized electron partial waves packets, in particular
in atomic problems~\cite{Hanetal:93a,ArgCol:09a}. Since B-splines are
non-orthogonal, however, they cannot be directly used in combination
with ATSP2K.  The expansion coefficients in Eq.~(\ref{pelB}) are
obtained through the following procedure.  First the MCHF orbitals,
computed on a radial grid using the \textsc{mchf} program of ATSP2K, are
fitted with B-splines using the \textsc{w\_bsw} program of the B-spline
R-matrix (BSR) package~\cite{Zat:06a}. The exterior complex scaling
radius, $R_0$, is chosen to be larger than the radius at which the MCHF
orbitals numerically vanish, therefore, the MCHF orbitals are expressed
in the region where $r$ is unscaled. Second, for each orbital obtained
at the MCHF stage, the B-spline corresponding to the largest mixing
coefficient in $P_{nl}(r)$ is removed. Finally, the remaining B-splines
are orthogonalized to the MCHF orbitals with the Gram-Schmidt algorithm.
When combined, the occupied $P_{nl}(r)$ and the orthogonalized B-spline
set form an orthonormal basis equivalent to the original B-spline radial
basis. The B-splines which do not vanish at $R_0$ are
complex~\cite{McCurdy2004b}; as a consequence, the final basis set is
complex as well.

\section{Atomic photoionization using exterior complex scaling}
\label{sec3}

The correlated many-electron basis set that we use to describe the
system before and after photo-absorption is obtained by diagonalizing
the full complex scaled Hamiltonian $H^\theta$ projected on the space spanned by the
ansatz in Eq.~(\ref{ciansatz}). The resulting
eigenenergies $E_k^{\theta}$, corresponding to the eigenvectors $\mid k^\theta\rangle$, are in general complex. Those
corresponding to bound states have vanishing imaginary part of the
energy; some complex eigenvalues, which correspond to resonances, are
largely independent of the rotation angle; finally, the remaining
eigenvalues correspond to the unstructured continuum. 

\subsection{The total cross-section}

The single-photon absorption cross-section can be expressed in terms of
the dynamic polarizability of the ground state $|0\rangle$ of the atom,
\beqa
\sigma(\omega) &=&-\frac{e^2}{4\pi \epsilon_0}\frac{4\pi}{3}\frac{\omega}{c}\nonumber\\
&&\hspace{0.4cm} \times \text{Im}
\la 0 \,| \,(\hat{\varepsilon} \cdot {\bf R})\, G_0^+(E_0+\hbar\omega)
\,(\hat{\varepsilon} \cdot {\bf R})\, |\, 0 \ra, \label{xsecCS}
\eeqa
where ${\bf R}=\sum_j{\bf r}_j$,
$G_0^+(E)=(E-H+i0^+)^{-1}$ is the Green's function of the field-free
hamiltonian,  $\hbar\omega$ is the photon energy, $\hat{\varepsilon}$ is the
direction of the laser polarization.

With complex scaling, the Green function is constructed as a
sum over the many-electron basis set components as~\cite{rescigno:75:cr,lindroth:95:liminus}

\beqa
 \sigma(\omega) &=& -\frac{e^2}{4\pi \epsilon_0}\frac{4\pi}{3}\frac{\omega}{c} \nonumber\\
&&  \times\ \text{Im}\left[ \sum_k
\frac{ \la 0\, | \,\hat{\varepsilon} \cdot {\bf R}^{(\theta)} \, |
k^{\theta} \ra \la k^{\theta} | \hat{\varepsilon} \cdot {\bf
R}^{(\theta)} | 0 \ra }{ E_0 -E_k^{\theta} + \hbar\omega}\right],
\label{xsecCS} 
\eeqa
 where $\mid k^{\theta} \rangle $ and  $\langle
k^{\theta}|$ are right and left eigenvectors  of the complex scaled
Hamiltonian $H^{\theta}$, with complex eigenvalues
$E^{\theta}_k$.  For the field-free Hamiltonian $H^\theta$, the left
eigenvectors are simply the transpose of the right
eigenvectors, see e.g.~\cite{bengtsson:08}.  The sum over $\mid k^{\theta}
\rangle$ goes over all eigenstates to $H^\theta$ including both
resonances and continuum states.  We note that since resonances appear
directly as eigenstates of $H^\theta$ the effect of a resonance on the
photoionization process is accordingly accounted for by a single term in
Eq.~(\ref{xsecCS}).

\subsection{The Fano profile}
 As was shown by Fano~\cite{Fan:61a} the
cross section in the vicinity of an isolated  resonance, situated  at
$E_r$ (relative to $E_0$) and with autoionization width $\Gamma$, can be
parametrized as 
\beq
 \label{Fano} \sigma(\epsilon)=
\frac{(q+\epsilon)^2}{1+\epsilon^2} \sigma_a + \sigma_b, 
\eeq
 where $q$
is the so-called asymmetry parameter and $\epsilon$ is defined as 
\beq
\label{epsilon} \epsilon = \frac{\hbar\omega-E_r}{\Gamma/2}, 
\eeq
 while
$\sigma_a$ and $\sigma_b$ are slowly varying as a function of $\epsilon$.

Looking at the  contribution from a {\em specific} resonance in the sum
on the right-hand side of Eq.~(\ref{xsecCS}), we can indeed put each such
contribution  in the form of Eq.~(\ref{Fano}).  The derivation is
detailed in Ref.~\cite{lindroth:95:liminus}, here we just give the
results.  Each resonance, $k$, contributes with 
\beq
 \label{sigmak}
\sigma_k(\epsilon_k)= \frac{(q_k+\epsilon_k)^2}{1+\epsilon^2} \sigma_a^k
+ \sigma_{min}^k, 
\eeq
 where the reduced energy $\epsilon_k$ of the
resonance $|k\rangle$ is defined as in Eq.~(\ref{epsilon}), with
$E_k^r=\text{Re}[E_k]-E_0$ being the position of the resonance, and
\mbox{$\Gamma_k = -2\text{Im}[E_k]$}.  The asymmetry parameter $q_k$,
but also $\sigma_a^k$ and $\sigma_{min}^k$, are here constants and are
determined by the resonance parameters and by the  complex dipole matrix
element between the ground state and the considered resonant state, the
nominator in the term in brackets in Eq.~(\ref{xsecCS}). Labelling it
$R_k+iI_k = \la k^{\theta} | {\hat{\varepsilon}} \cdot {\bf
R}^{(\theta)} | 0  \ra^2$~\cite{lindroth:95:liminus}, the
three constants are found to be \begin{eqnarray} \label{parameterq} q_k
& = &  b_k - \frac{I_k}{|I_k|} \sqrt{b_k^2+1} \\
\label{sigmaa} \sigma_a^k  & = &   \frac{e^2}{4\pi
\epsilon_0}\frac{4\pi}{3} \frac{1}{\hbar c} \left( \frac{R_k - 2 I_k
E_k^r/\Gamma_k}{2 q_k} \right)\\
\label{sigmamin} \sigma_{min}^k  & = &  -\frac{e^2}{4\pi
\epsilon_0}\frac{4\pi}{3} \frac{1}{\hbar c} \left( \frac{2 R_k
E_k^r/\Gamma_k +q_k^2 I_k }{2 q_k b_k} \right), \end{eqnarray} where

\beq
 b_k  =   \frac{\ I_k\ \Gamma_k  + 2 E_k^rR_k}{ R_k\Gamma_k - 2
E_k^r\ I_k }.  
\eeq
 If the sum over all the other contributions in
Eq.~(\ref{xsecCS}), from other resonances as well as from the smooth
continuum, just give slowly varying contributions in the vicinity of the
resonance $k$,  then it is clear that the total cross section is well
described by Eq.~(\ref{Fano}). This condition should be fulfilled if the
resonance  is far from other resonances and from thresholds. 

The resonance parameters, $E_r$, $\Gamma$ and  $q$ in
Eq.~(\ref{parameterq}) are often obtained from experiment through a fit
to a Fano profile. For validation purposes, we compare these with our
calculated values in Sec.~\ref{sec:res}.  Another parameter that is also
obtainable from experiments is the coefficient
$\rho^2$~\cite{FanCoo:65a}, 
\beq
 \rho^2 = \frac{\sigma_a}{\sigma_a +
\sigma_b}.  \label{rhosq} 
\eeq
 Here  $\sigma_a$ can be calculated
directly from Eq.~(\ref{sigmaa}), while $\sigma_b$  is not given by
$\sigma_{min}$ alone but from the full sum over states $k$ in
Eq.~(\ref{xsecCS}). However, in the limit that Eq.~(\ref{Fano}) is
valid, $\sigma_b$ can be obtained from Eq.~(\ref{xsecCS}) at
$\epsilon=-q$.


\subsection{Partial cross-sections}\label{sec:pwf}

\begin{figure} \caption{Radial perturbed function for the $3p
\rightarrow \epsilon d$ channel, in length gauge, for $\epsilon=-q$ in
the vicinity of the $3s 3p^6 4p$ resonance. The red and dashed blue
curves are its real and imaginary part, respectively.\label{cpwf}}
\includegraphics[width=\columnwidth]{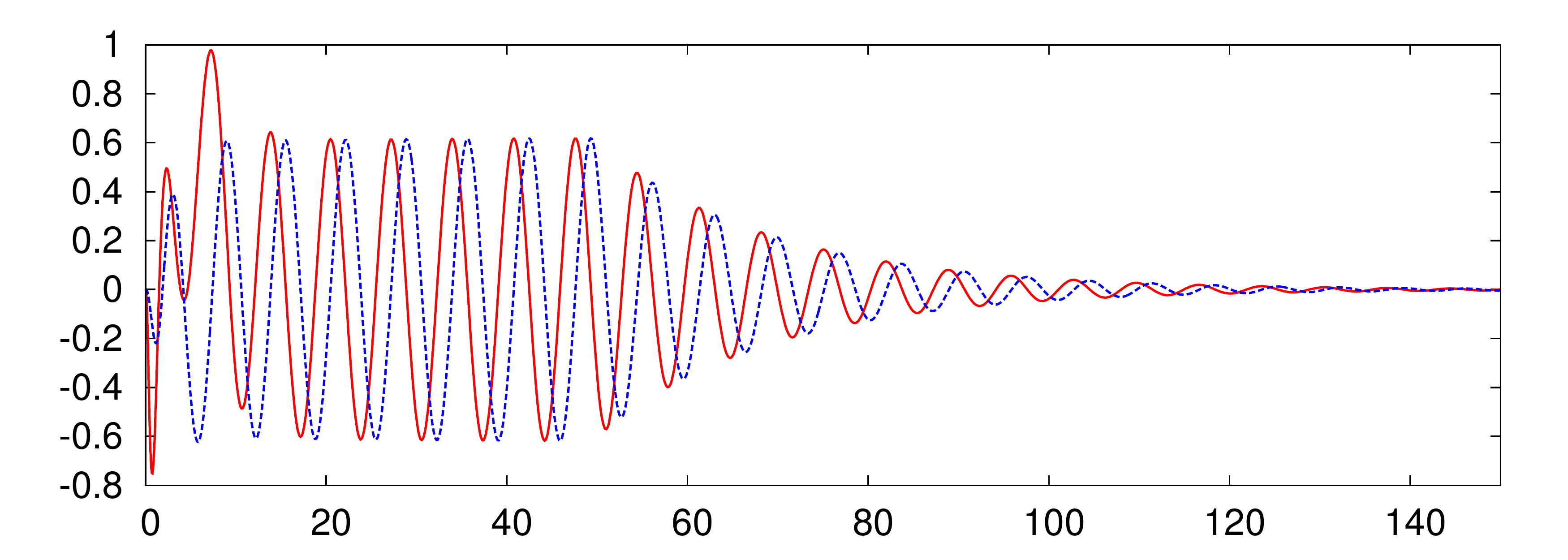} \end{figure}

Eq.~(\ref{xsecCS}) gives a convenient way to obtain the total cross
section. For the extraction of the partial cross sections we analyze the
many-electron, one-photon perturbed wave function, $\Xi(\omega,{\bf
r}_1,{\bf r}_2,\ldots ,{\bf r}_N)$, which is a function of the photon
energy as well as of the coordinates of all the particles in the system.
The perturbed wave function solves the following inhomogeneous
differential equation 
\beqa
 (H -E_0 - \hbar \omega)\Xi(\omega,{\bf
r}_1,{\bf r}_2,\ldots ,{\bf r}_{N+1}) &=&\nonumber\\
&&\hspace{-3.4cm} - \left({\hat{\varepsilon}} \cdot {\bf R}\right) \Psi_0({\bf r}_1,{\bf
r}_2,\ldots ,{\bf r}_{N+1})\, , 
\eeqa
 where the source term on the
right-hand side describes the excitation from the ground state by the
dipole operator acting on all electrons.  The concept of perturbed wave
functions was introduced by Sternheimer~\cite{sternheimer:1951}, and it
remains a useful tool also today, with novel applications emerging in
the field of attosecond science~\cite{marcus:tutorial:2012}.  Here  it
is obtained through direct summation over our many-body basis as 
\beq
\left|\Xi^\theta\right>= \sum_k \left| k^{\theta} \ra \frac{\la
k^{\theta} | {\hat{\varepsilon}} \cdot {\bf R}^{\theta}  | 0
\ra}{E_0-E_k^{\theta}  + \hbar \omega}\, , 
\eeq
 which represents the
full many-electron wave packet formed by absorption of one photon from
the initial state.  To extract partial
physical observables, $\Xi^\theta$ is projected on a specific channel, $\alpha$, defined by a bound $N$-electron state of the parent ion
and a fixed angular momentum of the outgoing electron. As a result, we
obtain a radial perturbed wave function (PWF), denoted
$\rho^\theta_\alpha(\omega,r)$, which is a superposition of the radial
functions $P_{nl}(r)$ introduced in Eq.~(\ref{ciansatz}) and
(\ref{pelB}).  Fig.~\ref{cpwf} shows the complex and imaginary parts of
$\rho^\theta_\alpha(\omega,r)$.  For $r<R_0$,
$\rho^\theta_\alpha(\omega,r)$ is equal to the radial perturbed wave
function without complex scaling, while in the complex scaled
region, $r>R_0$, it is artificially damped. As a second step, the
complex amplitude $A_\alpha(\omega)$ of the energy normalized PWF is
obtained by fitting $\rho_\alpha(\omega,r)$ to a linear combination of
the energy-normalized regular $F_l^E$ and irregular $G_l^E$ Coulomb
functions 
\beqa
 \text{Re}[\rho_\alpha(\omega,r)] &=&
\textrm{Im}[A_\alpha(\omega)] F^E_l(-1/k, kr) \nonumber\\ &&\hspace{1cm}
- \textrm{Re}[A_\alpha(\omega)] G^E_l(-1/k, kr) \label{eq:fitCoul},
\eeqa
 where, here, $k$ is the photoelectron wave number and $E$ is the corresponding energy.  The fit is carried
out for a radial distance $r<R_0$ so that it is not affected by the
complex scaling, but large enough that the asymptotic behavior is
established.
In practice, we compute $\Xi^\theta$, and $A_\alpha$, for a specific
 $M$ magnetic quantum number. In the following, however, observables extracted from the complex amplitudes are implicitly averaged over orientations except otherwise stated.
The absolute square of the amplitudes is related to the
electron flux and, hence, to the partial cross section, as

\beq
\label{pxsec} \sigma_\alpha(\omega) = \frac{4\pi}{3}\frac{e^2
}{(4\pi \varepsilon_0) \hbar c} a_0^{2}\ \frac{\hbar \omega}{\pi}
|A_\alpha(\omega) |^2.  
\eeq
 Note that the
 amplitude (and the cross section) is that of a  photoelectron in
 channel $\alpha$. The spin-averaged,
 angular dependent photoelectron amplitude, with given $M$ magnetic quantum number, is obtained after multiplying it by spherical harmonics coupled by Clebsch-Gordan coefficents
 as

\beq
A_{\alpha,M}(\omega) \sum_{m} Y_{lm}(\theta,\varphi)\ \langle  L_a (M-m); l
m | L_\alpha M \rangle \label{eq:angleA}, 
\eeq
 where $L_a$ is the angular momentum of the $N$-electron system $\Psi_a$, and  $L_\alpha$ that of the total
\mbox{$(N+1)$} - system. Expression~(\ref{eq:angleA}) can be used to obtain the
angular dependence of  photoelectrons ionized by polarized light 
\beq
\frac{d\sigma(\omega)}{d\Omega} = \frac{\sigma(\omega)}{4\pi} [ 1 +
\beta(\omega) P_2(\cos \phi)] 
\eeq
 where $P_2$ is the second order
Legendre polynomial and $\phi$ is the angle between the polarization
direction, and the direction of the photoelectron. The parameter
$\beta(\omega)$ is 
\beqa
 \beta(\omega) &=& \left(\sum_{\alpha\beta}
\Theta(\alpha;\beta) \text{Re}[A_\alpha^*(\omega) A_\beta(\omega)]
\right)\nonumber\\
&&\hspace{3cm}\times\ \left(\sum_{\alpha} |A_\alpha(\omega)|^2 \right)^{-1}.
\eeqa
 where $\Theta(\alpha;\beta)$ is a purely angular factor given by
 \beqa
\Theta(\alpha;\beta) &=&
(-1)^{L_0-L_a+l_\alpha+l_\beta+1}\ i^{-l_\alpha+l_\beta} \delta(a,b)\nonumber\\
&&\times\ \sqrt{\frac{2 (2l_\alpha+1)(2l_\beta+1)(2L_\alpha+1)(2L_\beta+1)}{3}} 
\nonumber\\
&&\times\ \langle l_\alpha 0 l_\beta 0 | 2 0 \rangle 
\left \{\ar{ccc}
l_\alpha & l_\beta & 2 \\
L_\beta & L_\alpha & L_a 
\ear\right\}
\left \{\ar{ccc}
1 & L_\alpha & L_0 \\
L_\beta & 1 & 2 
\ear\right\}
\nonumber\\
 \eeqa
 where the index $0$ refers to the initial state, and $\alpha=(a,l_\alpha)$ and $\beta=(b,l_\beta)$ 
refer to the two interfering channels.

\subsection{Extraction of resonance properties from the partial cross section}
\label{subsec:extract}
Properties like $\rho^2$, defined in Eq.~(\ref{rhosq}), can be
extracted from the partial channels as well as from the total cross
section. For this we follow  Kabachnik and  Sazhina~\cite{KabSaz:76a}
and Sorensen~\etal~\cite{Soretal:94a}, and express the amplitude
of the PWF as the combination of a \emph{resonant}  $A^R_{\alpha}$ and
a \emph{decoupled}  $A^D_{\alpha}$ complex amplitude 
\beqa
A_{\alpha}(\epsilon)&=& \mathcal{F}(\epsilon) A^{R}_{\alpha} +
A^D_{\alpha}(\epsilon)\label{Atraj}\\
\mathcal{F}(\epsilon)&=&\frac{q+\epsilon}{i+\epsilon} = (\epsilon-i)
\frac{q+\epsilon}{1+\epsilon^2} \label{Ccirc}, 
\eeqa
 where $\epsilon$ is
defined as in Eq.~(\ref{epsilon}). 
If the decay of the bound component of the resonance is treated to lowest order
in the Coulomb interaction, these amplitudes have the same phase up to a $\pi$ shift~\cite{FanCoo:65a,KabSaz:76a}. We show in Section~\ref{subsec:pxsec} that it remains valid in the case of the $3s3p^6np$ resonances of Ar, and that the effect of the higher order contributions, present in our approach, is small.
From $A^R_{\alpha}$ and
$A^D_{\alpha}( \epsilon)$ , we can calculate $\rho^2$~\cite{FanCoo:65a}
as 
\beq
 \rho^2= \frac{\sigma_a}{\sigma_a+\sigma_b} = 1 -
\frac{\sigma_b}{\sigma_a+\sigma_b} = 1 - \sum_\alpha
\frac{|A^D_{\alpha}|^2}{|A^{R}_{\alpha} + A^D_{\alpha}|^2}, \label{rho}
\eeq
 where we have used the fact that the non-resonant background cross section,
$\sigma_b$, is obtained from the sum of the squares of the decoupled
amplitudes, $A^D_{\alpha}$.
%
%
For the extraction of $A^R_{\alpha}$ and $A^D_{\alpha}( \epsilon)$, we
suppose first that $A^{R}_{\alpha}$ can be considered a constant.
Provided $A^D_{\alpha}(\epsilon)$ is slowly varying, we
are able  to describe it  around a certain $\epsilon$ as a
$n$-term Taylor expansion.  Then $A^R_\alpha$ can be obtained from
$n^{th}$ derivative of Eq.~(\ref{Atraj}) as 
\beq
\label{Ar} A^R_\alpha =
\frac{d^n A_\alpha(\epsilon)}{d\epsilon^n} \left(\frac{d^n
\mathcal{F}(\epsilon)}{d\epsilon^n}\right)^{-1}\, .  
\eeq
 The number of
terms, $n$, used is decided from the convergence of $A^R_\alpha$ for a
given $\epsilon$, and the stability of  $A^R_\alpha$ when $\epsilon$ is
varied can be used for assessing the reliability of the procedure.
Making a comparison between the values obtained for $\rho^2$ from the total
cross section Eq.~(\ref{rhosq}), and from the partial amplitudes, the
right-hand side of  Eq.~(\ref{rho}), is a test of the extraction of the
partial cross sections as well as of the Fano parametrization itself.



\section{Application to Argon $3s\rightarrow np$ resonance region}\label{sec:res}

\subsection{Details of the calculations}

In the present section, we briefly outline the construction of the MCHF
model. Our approach is similar to those followed by Berrah~\etal~\cite{Beretal:96a}
and by Burke and Taylor~\cite{BurTay:75a}; the
reader can find further details in the original references.  The
spectroscopic orbitals $\{1s,2s,2p,3s,3p\}$ are optimized by performing
a Hartree-Fock (HF) calculation on the $3p^5~^2P^o$ state of Ar$^+$.
Then, the $\{\overline{3d}, \overline{4s}, \overline{4p}\}$ correlation
orbitals are optimized in a MCHF calculation on the lowest $^2S$ state
of Ar$^+$ including single excitations of the $3s3p^6$ configuration, as
well as the $3s3p^4 \overline{3d}^2$ complex.
As illustrated in Table~\ref{mix2S},
this state is characterized by
strong configuration interaction.

\begin{table} \caption{List of mixing coefficients in the MCHF expansion
of the lowest $^2 S$ $\left( 3s 3p^6 \right)$ state of Ar$^+$.
\label{mix2S}} \begin{ruledtabular} \begin{tabular}{lr}
Configuration state function (CSF) & mixing coefficient \\ \hline $3s
3p^6$               &   0.83498723 \\ $3s^23p^4(^1D)3d$        &
-0.52070520 \\ $3s  3p^4(^1S)3d^2(^1S)$ &    0.09817089 \\ $3s
3p^4(^3P)[^4P]3d^2(^3P)$  & -0.09655511 \\ $3s  3p^4(^1D)3d^2(^1D)$ &
-0.08964443 \\ $3s  3p^5[^1P]4p       $ &  -0.05536407 \\
$3s^23p^4(^1S)4s       $ & -0.03427725 \\ $3s  3p^4(^3P)[^2P]3d^2(^3P)$
& -0.01888098 \\ $    3p^6 4s$                 &  0.00827551 \\ $3s
3p^5[^3P]4p       $  & -0.00088679 \\
\end{tabular} \end{ruledtabular} \end{table}

The set of CSF included in the diagonalization of the $N$-electron
Hamiltonian defining the $^2S$ parent states entering the close-coupling expansion (see
Eq.~(\ref{ciansatz})), is the same as the one used in the MCHF
calculation. For the $^2P^o$ parents, we include the single
excitations of the $3s^23p^5$, and the doubly excited CSF arising from the
$3s^23p^3\overline{3d}^2$, $3p^5\overline{3d}^2$ and $3p^6\overline{4p}$
configurations.

This model yields a threshold energy for the $3s$ photoionization
of Argon of $29.73$~eV, compared to the experimental value of
$29.24$~eV~\cite{Kiketal:96a}. The difference of $0.49$~eV between
theory and experiment is due to the lack of balance between the
description of the ion and of the ground state of the neutral atom.
Since this discrepancy is mostly reflected in a collective shift 
of all spectral features, we simply present all
our results on a photon energy scale shifted downward by $0.49$~eV to
compare with the experiment, rather than using any semi-empirical
corrections.

The set of localized configuration state functions $\chi_\xi$
in Eq.~(\ref{ciansatz}) comprises all single and double excitations from
$3s^23p^6$, all triple excitations with at least two $\overline{3d}$
electrons, and the $3s^23p^2\overline{3d}^3 \overline{4p}$
configuration, i.e., all those listed in Table~2 of
Ref.~\cite{Beretal:96a}. 
 
The final close-coupling expansion using B-splines is obtained as
explained in Section~\ref{close-coupling ansatz}-\ref{bsplines}.  For
small radii, a dense B-spline basis is necessary to describe correctly
the inner orbitals. 
A $0.1/Z~a_0$ step size is used from the origin to $1/Z~a_0$. Then, the
knot spacing increases exponentially up to $0.9~a_0$. The knot sequence
is then linear up to $80~a_0$. It is generated by the
\texttt{W\_BSW} program of the BSR package~\cite{Zat:06a}.  To represent
high-$n$ Rydberg states, we extend the node set with an exponential
sequence of 100 nodes from $80$ to $R=400~a_0$. The final basis contains
249 B-splines.  We ascertain that the results are stable with respect to
changes of the complex scaling parameters by varying $R_0$, the radial
distance at which exterior complex scaling begins, from 50 to $80~a_0$,
and by doubling the complex scaling angle, $\theta$ in Eq.(\ref{exterior}).

\subsection{Total photoionization cross-section and  Fano profile
parameters}

\begin{figure} \caption{Top panel: Details of the complex energies of
the field-free hamiltonian of argon calculated for the $^1P^o$ symmetry
with $\theta=0.055$~rad (red circles) and $0.11$~rad (black squares).
Bottom panel: Comparison of the total calculated cross sections (thick
lines), in length (full red curve) and velocity (dashed blue curve)
gauges. The black dots are the experimental values of Ref.~\cite{Beretal:96a}, in arbitrary units, scaled as in the original paper. The thin lines are the cross
sections computed by including only the $3s 3p^6 np$ resonances in the
basis set. \label{resfig}} \includegraphics[width=\columnwidth]{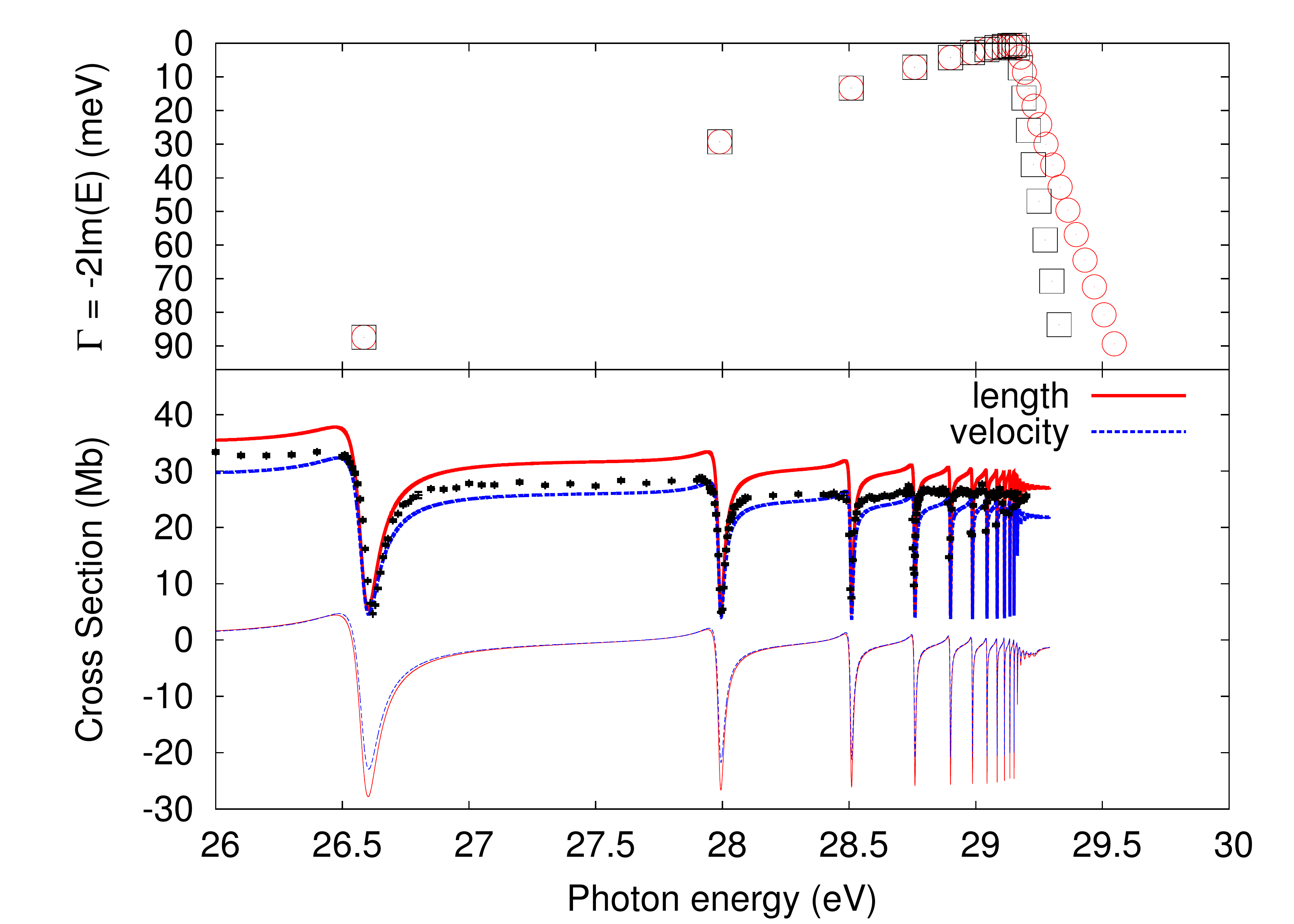}
\end{figure}

The upper panel of Fig.~\ref{resfig} shows part of the complex rotated
energy spectrum for the $^1P^o$ manifold, obtained from the
diagonalization of the field-free Hamiltonian $H^\theta$ using two
different complex rotation angles.  As expected, the complex resonance
energies are virtually independent of the rotation angle. For example,
all the parameters of the $3s3p^{6}np$ resonances agree to one ppm up to
the $n=13$ state. The pseudo-continuum energies, instead, follow a
$\theta$-dependent path in the complex plane~\cite{bengtsson:08}. Also
Rydberg states with $n>13$, that do not fit in the computational box,
show a $\theta$-dependence.

The lower panel of Fig.~\ref{resfig} shows total cross-sections obtained
in length and velocity gauge. The thick lines correspond to the
cross-sections obtained by including the complete sum over the basis set in
Eq.~(\ref{xsecCS}), while the thin lines are computed including only the
$3s 3p^6 np$ resonance series by summation over the corresponding $\sigma_k$ given in Eq.~(\ref{sigmak}). The
latter curves reproduce well the resonant structure of the cross-section,
but the remaining non-resonant contributions are required to
obtain the absolute value of the cross-section. 
The good agreement of the calculated total cross-section in the two gauges in the transmission window suggests that $\sigma_b$ in Eq.~(\ref{Fano}) is well described and that the discrepancy comes mostly from $\sigma_a$.

\begin{table*} \begin{ruledtabular} \caption{ Comparison of the
resonance parameters calculated in this work with the ones
fitted to R-matrix and experimental cross-sections for the $3s 3p^6 np$
resonances.  For each resonance, parameters computed in length gauge are
given on the first line, and the ones computed in velocity gauge on the
second line.
\label{restab}}
\begin{tabular}{lD{.}{.}{3}D{.}{.}{1}D{.}{.}{3}D{.}{.}{4}cD{.}{.}{3}D{.}{.}{1}D{.}{.}{3}D{.}{.}{3}cD{.}{.}{3}cD{.}{.}{3}D{.}{.}{6}D{.}{.}{6}}
&  \multicolumn{4}{c}{ this work} && \multicolumn{4}{c}{
	R-matrix~\cite{Beretal:96a}}&&  \multicolumn{1}{c}{
		Exp.~\cite{Wuetal:95a}  } && \multicolumn{3}{c}{
			Exp.~\cite{Beretal:96a} } \\
			\cline{2-5}\cline{7-10}\cline{12-12}\cline{14-16}
			$n$ &
			\multicolumn{1}{c}{$\Delta E$} & \multicolumn{1}{c}{$\Gamma$} &
			\multicolumn{1}{c}{$q$} & \multicolumn{1}{c}{$\rho^2$ }  &&
			\multicolumn{1}{c}{$\Delta E$} & \multicolumn{1}{c}{$\Gamma$} &
			\multicolumn{1}{c}{$q$ }& \multicolumn{1}{c}{$\rho^2$ }&&
			\multicolumn{1}{c}{$\Delta E$} && \multicolumn{1}{c}{$\Gamma$} &
			\multicolumn{1}{c}{$q$ }& \multicolumn{1}{c}{$\rho^2$ }\\
& \multicolumn{1}{c}{eV} & \multicolumn{1}{c}{meV} &  &  & &
\multicolumn{1}{c}{eV} & \multicolumn{1}{c}{meV} & & & &
\multicolumn{1}{c}{eV} && \multicolumn{1}{c}{meV} \\
 \hline $4$ & 26.585
& 87.4 & -0.389 & 0.8391 && 26.633 & 83.8 & -0.383 & 0.843 && 26.605 &
&80.2(7) & -0.286(4) & 0.840(3)\\
 &  & & -0.442 & 0.8350 &&            &
& -0.433 & 0.846\\
 $5$ & 27.990 & 29.3 & -0.299 & 0.8452 && 27.997 & 27.4 & -0.292 & 0.825 && 27.994 &
& 28.5(8) & -0.177(3) & 0.848(3)\\
 &  &
& -0.345 & 0.8411 && &             & -0.342 & 0.829\\
 $6$ & 28.509 &
13.3 & -0.266 & 0.8467 && 28.508 & 12.4 & -0.262 & 0.824 && 28.509 &
&12.2(3) & -0.135(9) & 0.852(9)\\
 &  &   & -0.310 & 0.8425 && &
& -0.312 & 0.827\\
 $7$ & 28.760 &  7.1 & -0.250 & 0.8475 && 28.756 & 6.7 & -0.249 & 0.823 && 28.757 &
 &  6.6(1) & -0.125(4) & 0.846(9)\\
 &  &
& -0.293 & 0.8433 && &             & -0.299 & 0.827\\
 $8$ & 28.901 &
4.3 & -0.241 & 0.8476 && 28.896 &  4.0 & -0.240 & 0.824 && 28.898 &
& 4.5(2) & -0.132(4) & 0.77(2)\\
 &  &   & -0.284 & 0.8434 && &
& -0.291 & 0.827 \\
 $9$ & 28.987 &  2.8 & -0.236 & 0.8477 && 28.928 & 2.6 & -0.235 & 0.823 &&        & 
& 4.1(2) & -0.115(8) & 0.63(3)\\
 &  &
& -0.279 & 0.8435 && &             & -0.286 & 0.826\\
 $10$& 29.045 & 1.9 & -0.233 & 0.8479\\
 &  & & -0.275 & 0.8436\\
 $11$& 29.085 &  1.3 &
-0.230 & 0.8480\\
 &  & & -0.272 & 0.8437\\
 $12$& 29.114 &  1.0 & -0.228
& 0.8480\\
 &  & & -0.271 & 0.8438\\
 $13$& 29.135 &  0.7 & -0.227 &
0.8481\\
 &  & & -0.269 & 0.8438
\end{tabular} \end{ruledtabular}
\end{table*}

Table~\ref{restab} lists the parameters for some of the $3s 3p^6 np$
autoionizing states computed with our method, in length and velocity
gauges, and compares them to Ref.~\cite{Beretal:96a,Wuetal:95a}. Our
results are generally consistent with those obtained with the R-matrix
method~\cite{Beretal:96a} with similar choices for the computational
model and basis. The widths are slightly larger than
the experimental ones and those obtained from the fits to the R-matrix
calculation. The agreement of the $q$-parameter from the R-matrix
calculation is good, but both calculations show rather large
deviations from the experiment. Ref.~\cite{Beretal:96a} attributes this
discrepancy to the difficulty to accurately represent the
non-resonant background in the close-coupling expansion. 
The marked gauge dependence of $\sigma_a$ discussed in the previous paragraph supports the fact than the contribution from the resonant channel
is not yet converged.

\subsection{Partial cross sections and angular distribution}
\label{subsec:pxsec}

\begin{figure} \caption{Top panel: partial cross-sections for $3p$
ionization and total cross-section, in length gauge. Bottom panel:
relative error of the sum of the above partial cross-sections and the
total cross-section computed with Eq.~(\ref{xsecCS}), for $R_0=65~a_0$
and $\theta_0=0.055$~rad.\label{partxsec}}
\includegraphics[width=\columnwidth]{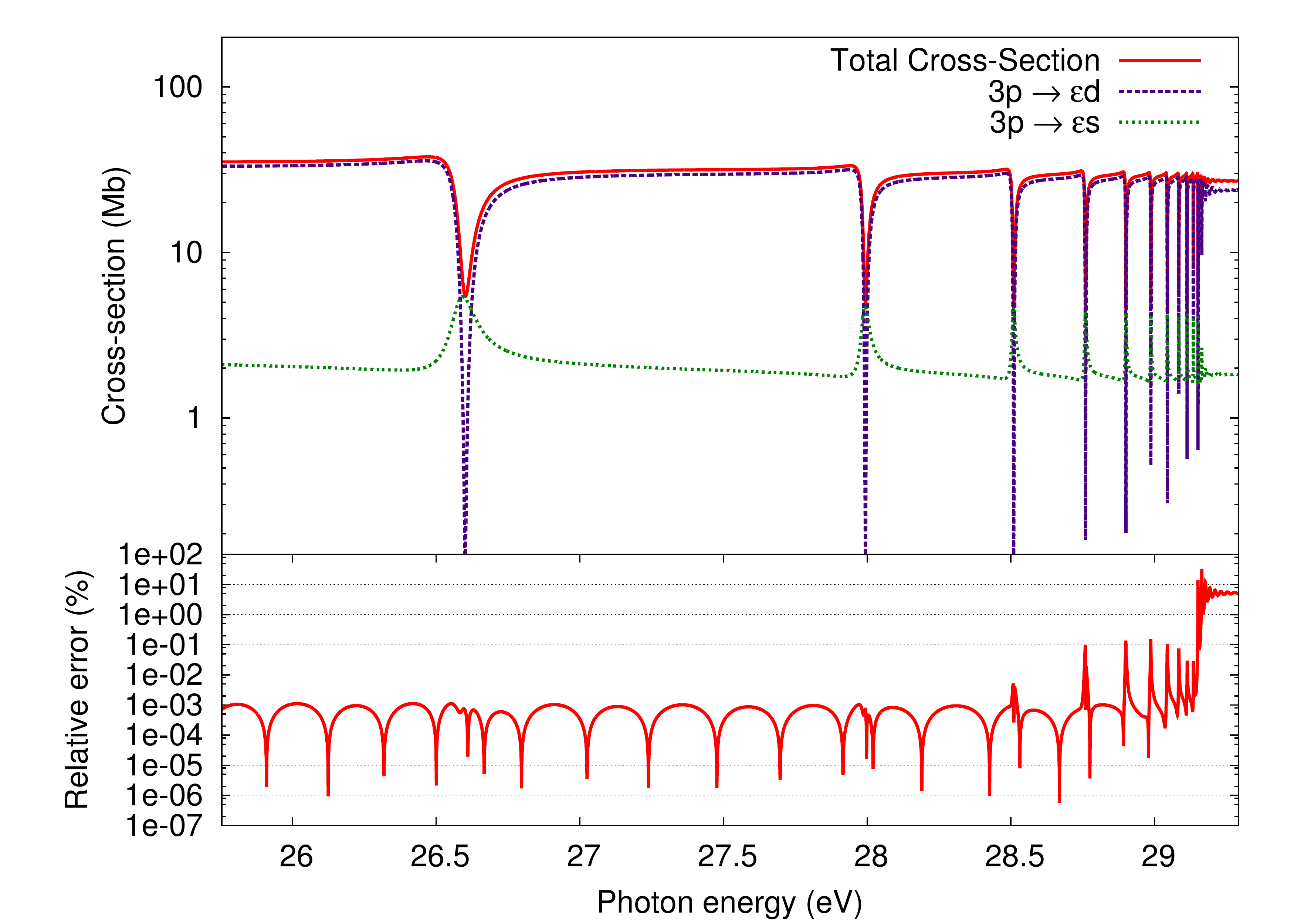} \end{figure}

As a test of the robustness of the method outlined in
Sec.~\ref{sec:pwf}, we compare the total cross-section computed with
Eq.~(\ref{xsecCS}) to the sum over the partial amplitudes of the $3p$
photoemission channels computed with Eq.~(\ref{pxsec}). This is shown for the length gauge in Fig.~\ref{partxsec}. As can be
seen in the lower panel, the relative error between the two approaches
is of the order of 10~ppm at most energies, which is close to our
computational accuracy. For the high $n$ resonances, spurious long range
effects affect the channel perturbed radial functions. Therefore, their
matching to Coulomb waves, see Eq.~(\ref{eq:fitCoul}), is less accurate at
the considered $R_0$, but the error remains consistently of the order of
0.1\%. Above the $3s$ ionization threshold, the $3s3p^6\epsilon p$
channel accounts for the 10\% discrepancy that can be seen in the
for photon energies above $29.24$~eV.

\begin{figure} \caption{Norm, in atomic units, and argument of the partial photoionization
complex amplitudes versus the reduced energy of the $3s 3p^6 4p$
resonance. (a) $3p \rightarrow \epsilon d$ channel. (b) $3p \rightarrow
\epsilon s$ channel. The crosses show 1 on 7 computed values. The lines are
obtained from Eq.~(\ref{eq:an}). Red and blue indicates length or
velocity gauge, respectively.\label{phases}}
\includegraphics[width=\columnwidth]{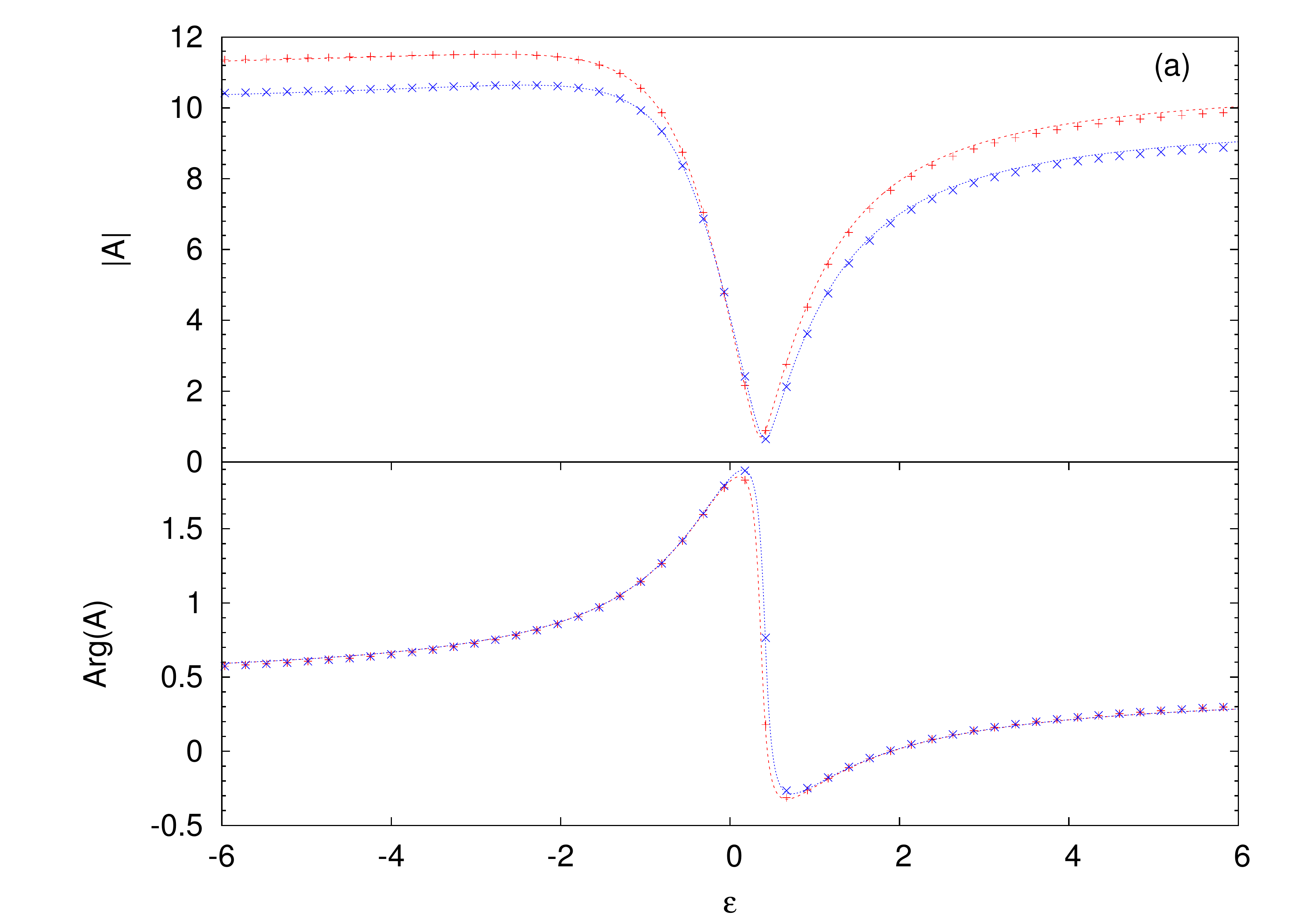}
\includegraphics[width=\columnwidth]{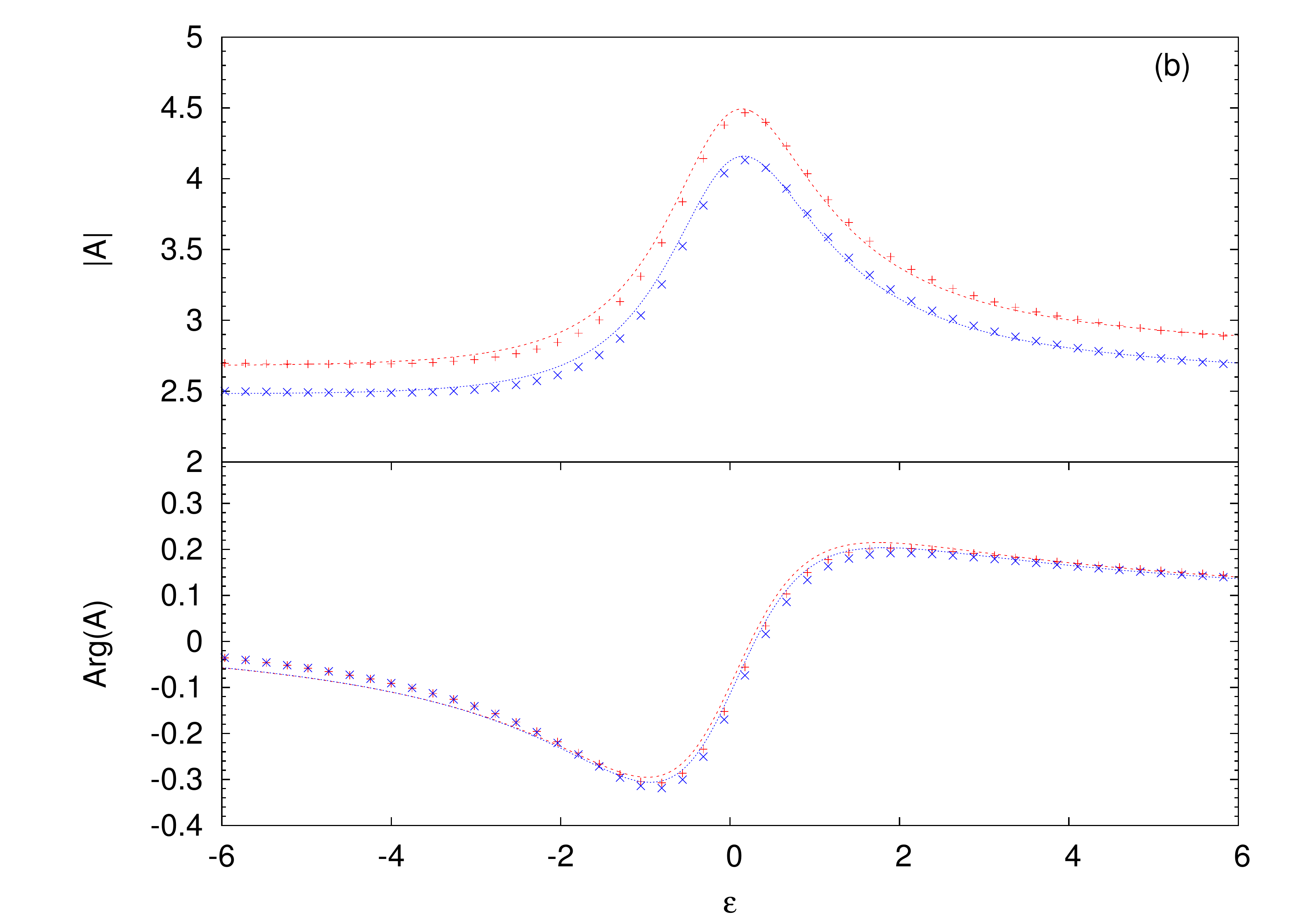} \end{figure}

We show in Fig.~\ref{partxsec} that the window-shape of the resonances
comes from the dominating $3p\rightarrow\epsilon d$ channel, while the
$3p\rightarrow\epsilon s$ cross-section is instead enhanced on resonance. 
The partial amplitudes
cannot generally be described by the same set of resonance shape parameters.
Here  $\sigma_{ s}$ has,
for example,  its maximum value at $\epsilon=-q \approx 0.4$, where the total cross
	section has a minimum.  This is due to a destructive interference
	between $A_{ s}^R$ and $A_{ s}^D$ that suppresses the partial
	background amplitude. This is illustrated in
	Fig.~\ref{phases} where the markers correspond to the norm and argument of
	the calculated partial complex amplitudes in the vicinity of the $3s
	3p^6 4p$ resonance, while the curves are obtained
	from the simplified expression
\beq
\label{eq:an} \frac{q+\epsilon}{i+\epsilon} |A_\alpha^R| + \eta
|A_\alpha^D(0)| 
\eeq
where $\eta=1$ for $\alpha=d$, and $\eta=-1$ for $\alpha= s$. Here, we drop the $3p^5$ label in $\alpha$ for conciseness. The good
agreement between the curves and data points illustrates the reliability of the
extraction of $A_\alpha^D(\epsilon)$ and $A_\alpha^R$ and that
$A_\alpha^D(\epsilon)$ is a slowly varying function. More interestingly, it means that $A_{
d}^D(\epsilon)$ and $A_{ d}^R$ are in phase $(\text{Arg}[A_{
d}^D(\epsilon)/A_{ d}^R ] \approx 0)$ while $A_s^D(\epsilon)$ and
$A_s^R$ are in phase opposition $(\text{Arg}[A_{ s}^D(\epsilon)/A_{
s}^R ] \approx \pi)$ over a wide range of $\epsilon$. As mentioned in Section~\ref{subsec:extract}, these 
relations are rigorously valid if the decay of the bound component of the resonance is treated to lowest order
in the Coulomb interaction~\cite{FanCoo:65a,KabSaz:76a}. However, while
they hold with increased accuracy as the width of the resonance
decreases, they are not exactly realized within more general
models like ours.

Another test of our procedure for extracting the $A_\alpha^D(\epsilon)$
and $A_\alpha^R$ values, and in particular of $A_\alpha(\epsilon)$, is
the calculation of $\rho^2$. The values from the right-hand side of
Eq.~(\ref{rho}) agree with those obtained from the total cross section
as explained in Sec.~\ref{sec2} within a couple of parts per thousand.
We compare the values obtained from Eq.~(\ref{rho}) to the fitted ones
from Berrah~\etal~\cite{Beretal:96a} in Table~\ref{restab}. The present
results in length gauge agree with experiment up to the $3s7p$ state,
while the velocity gauge underestimates $\rho^2$ slightly.
The R-matrix fitted values are systematically
below ours and the experiment, except in the case of the $3s 3p^6 4p$ resonance.

\begin{figure} \caption{Calculated beta parameter ($\beta$) in length
(full red line) and velocity (broken blue line) compared to
experiment~\cite{Beretal:96a} (black dots). The vertical dashed line
indicates a change of photon energy scale.
\label{beta}} \includegraphics[width=\columnwidth]{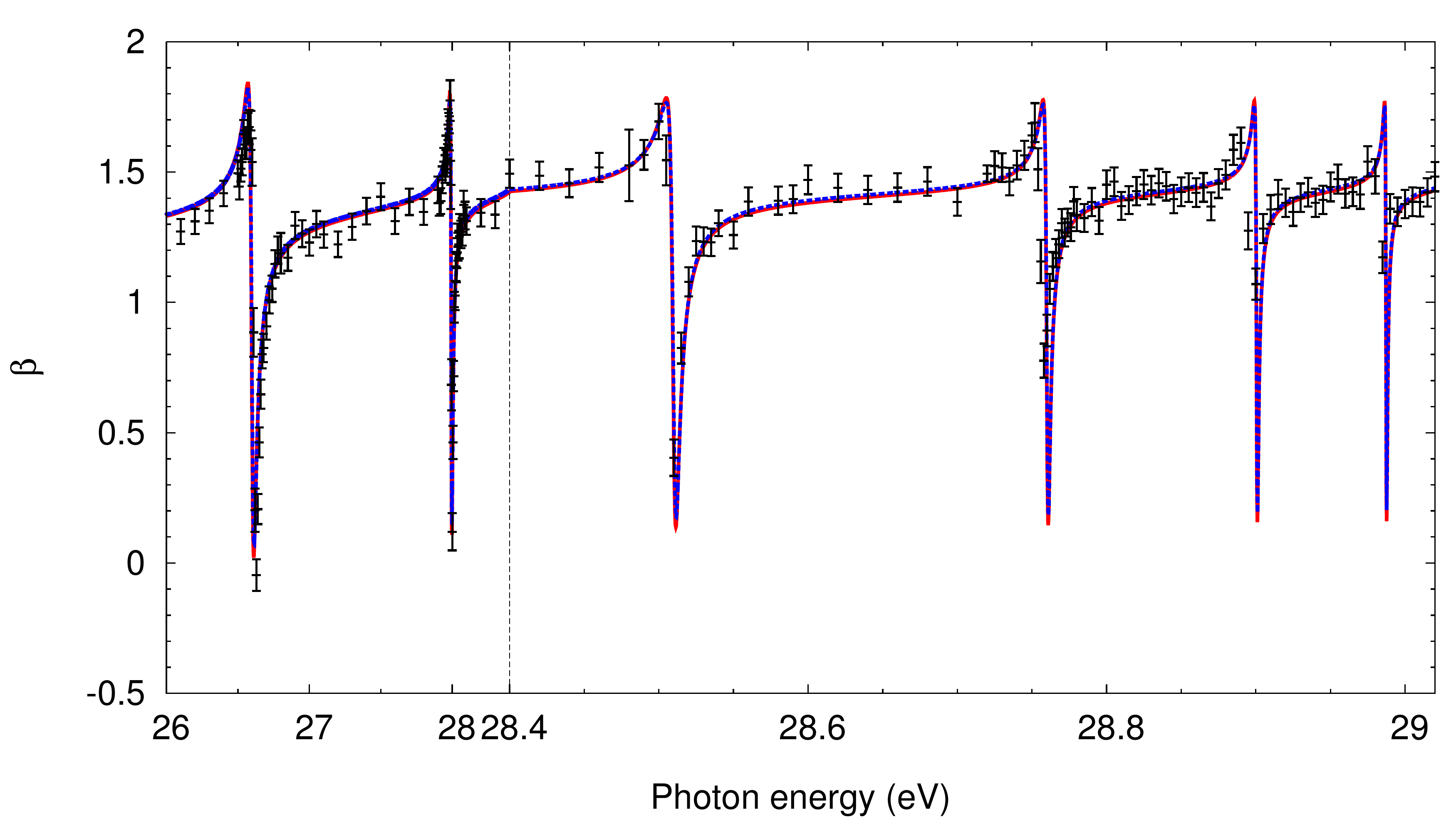} \end{figure}

Finally, Fig.~\ref{beta} compares the calculated $\beta(\omega)$ for the
$3snp$ resonances, $n=4$ to $9$, to the experimental data of
Berrah~\etal~\cite{Beretal:96a}. Overall, the agreement is excellent.
No convolution is applied on the theoretical results so that discrepancies
appear close to the high $n$ resonances. The $\beta$ parameter is, as expected, about $1.5$ where the $d$ channel dominates,  far from the resonances  (see Fig.~\ref{partxsec}), while the
photoelectron angular distribution is anisotropic ($\beta=0$) when the
$s$ channel dominates, \ie in the vicinity of the minimum of each Fano structure.  The
structure in between those two extremes is well described in the
calculations, which means that, despite the difference between theory
and experiment regarding $q$-values, the relative amplitude and phase
between the $s$ and $d$ channels, leading to angular dependent
interferences, is well represented.

\section{Application to photoionization delay measurements}
\label{sec:delay}

Photoionization by attosecond XUV pulses assisted by phase-locked IR
laser-probe fields  is a widely used experimental scheme to gain
temporal information about subfemtosecond pulse structures and dynamics
\cite{FerencIvanov2009}.  This laser-assisted photoionization process is
often interpreted as an attosecond streak-camera method, where the
photoelectron will gain momentum from the instantaneous vector potential
of the probe field depending on its exact time of ionization by the
attosecond pulse.  Assuming that the intensity of both pulses is weak,
the corresponding light-matter interaction can be treated using
lowest-order perturbation theory and the {\it on-set} of this streaking
mechanism can be described with high fidelity using only a limited
number of complex transition amplitudes \cite{Dahetal:12a}.  The
delay-dependent probability modulations of the photoelectron
distributions are then directly related to the phase differences of the
two-photon, two-color matrix elements (XUV$\pm$IR). These so-called
\emph{atomic phases} in laser-assisted photoionization
\cite{TomMul:02a}, have been implemented for reliable characterization
schemes of attosecond pulses \cite{Paul:2001}.  Recently, similar
experimental techniques have been extended to study the delay in
photoionization from different initial states in neon
\cite{Schultze2010a} and argon atoms \cite{klunder:2011,Gueetal:12a}.
Theoretical work has shown that the total atomic delay, $\tau_A$,
measured in these experiments can be approximated as the sum of the
one-photon Wigner-like delay in photoionization, $\tau_W$, plus the
so-called continuum--continuum delay, $\tau_{cc}$, which is induced by
the laser-probe field and the long-range ionic potential, so that
\cite{klunder:2011,Nagele:2011,Dahetal:12a} 
\beq
\tau_A=\tau_W+\tau_{cc}\ .  \label{simpledelays} 
\eeq

In order to evaluate $\tau_W$,
we first apply the method presented in
Sec.~\ref{sec:pwf} in order to extract the partial wave amplitudes,
$A_\alpha(\omega)$ which contain the asymptotic phase-shifts,
$\delta_\alpha(\omega)\equiv\text{Arg}[A_\alpha(\omega)]$, relative to
the Coulombic phase-shifts, $\sigma_{l}(\omega)$, where $l$ is the
angular momentum of the photoelectron in channel $\alpha$. Then, using
the total asymptotic phase shift, $\eta_\alpha(\omega)\equiv
\sigma_{l}(\omega)+ \delta_\alpha(\omega)$, we construct the Wigner-like
delay for channel $\alpha$ as a finite-difference derivative 
\beq
\tau_W^{(\alpha)} \equiv
\frac{\eta_\alpha(\omega_>)-\eta_\alpha(\omega_<)}{2\omega_{IR}},
\label{Wignerlikedelay} 
\eeq
 where the difference between the XUV
frequencies that will interfere due to the influence of the probe field
are separated by two IR probe photons, $\omega_>-\omega_<=2\omega_{IR}$.

 The absorption of a laser-probe photon, required to make the
transition from the intermediate momentum $k_i$ to the final momentum
$k_f$, induces a phase-shift denoted $\phi_{cc}$. Starting from the formalism presented in
Ref.~\cite{marcus:tutorial:2012}, we use the soft-photon approximation,
$ k_f^2/2\gg \omega_{IR}$, and find the following expression for the
phase-shift, using atomic units ($\hbar=1$, $e=1$,
$m_e=1$, $4\pi\epsilon_0=1$), 
\beq
 \phi_{cc}(k_f,\omega_{IR})= \arg \left\{
	\left(\frac{k_f}{i\omega_{IR}}\right)^{i\omega_{IR} Z/k_f^3} \Gamma
	\left[ 1+\frac{i\omega_{IR} Z}{k_f^3} \right] \right\},
	\label{softphotonccphase} 
\eeq
 where $Z=1$ is the charge of the target ion.
	Eq.~(\ref{softphotonccphase}) is accurate for photoelectrons with high
	kinetic energy.
The continuum-continuum  delay can then be written as 
\beq
\tau_{cc}(k_f,\omega_{IR})\approx
-\frac{\phi_{cc}(k_f,\omega_{IR})}{\omega_{IR}}\, .
\label{softphotonccdelay} 
\eeq
Alternatively this delay could be estimated with other methods such as 
the eikonal Volkov approximation~\cite{IvaSmi:11a} or 
classical-trajectory Monte-Carlo simulations~\cite{Nagetal:11a}.

If there is more than one channel contributing to a given final state,
the two-photon amplitudes have to be added in a consistent way depending
on the experimental observable \cite{TomMul:02a}. Such a
detailed analysis is beyond the scope of this paper, where we will show that
the correlated Wigner-like delay Eq.~(\ref{Wignerlikedelay}) from the dominant channels is enough
to explain the observations within the experimental uncertainty.






\begin{figure} \caption{\label{fig:3s3pdel} (a) Wigner-like delays for
photoionization from argon.  There is no perceptible gauge dependence for
photoionization from the $3p$ orbital. For the
$3s$ photoionization, the  results from velocity (blue broken) and length gauge (red dashed) are shown.  (b) Wigner-like delay difference between photoionization from the 
$3p$ and $3s$ orbitals in argon.  The thin curves are obtained with the
full model that exhibit a pseudo-resonance at 39.4~eV.  The thick curves
are obtained with the truncated model for velocity (blue dashed) and
length (red full) gauge, respectively. The experimental values of Ref.~\cite{Gueetal:12a} are given with error bars. The experimental data point obtained using the $\tau_{cc}$ from Eq.~(\ref{softphotonccdelay}) and Ref.~\cite{IvaSmi:11a} are given by the circles and squares, respectively.}
\includegraphics[width=\columnwidth]{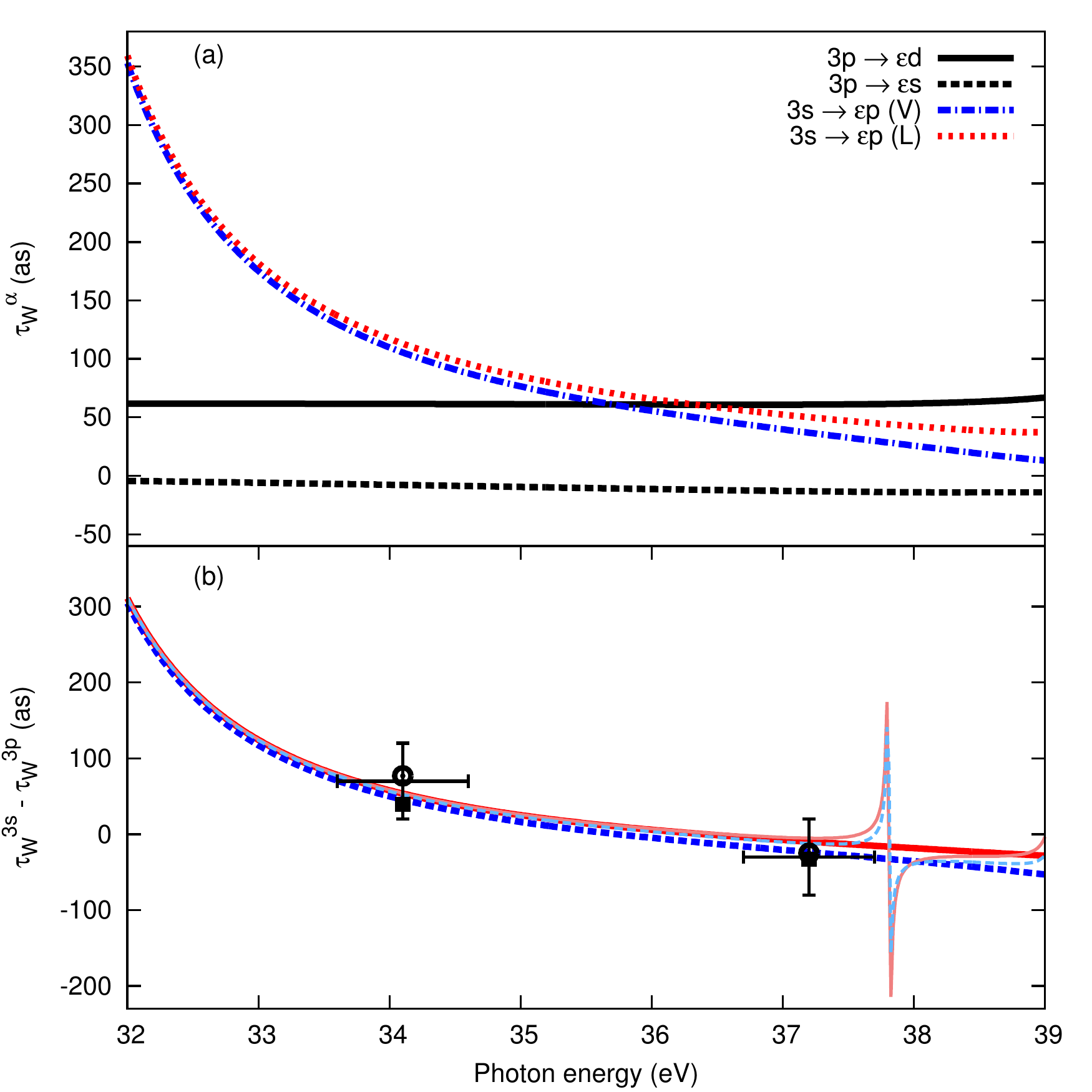} \end{figure}

\medskip

Kl\"under~\etal~\cite{klunder:2011} and
Gu\'enot~\etal~\cite{Gueetal:12a} have measured the $3s-3p$
atomic delay difference at XUV photon energies of $34.1$, $37.2$ and
$40.3$~eV corresponding to the high-order harmonic sidebands 22, 24 and 26 for a generating field of $1.55$~eV.  The contribution from the continuum--continuum delay was
then subtracted according to Eq.~(\ref{simpledelays}), in order to
isolate the Wigner-like delay in photoionization.

The first pseudo-resonance in our model lies at an energy of 39.4~eV, which is inside the range of the experimental measurement, leading to an unphysical feature at $37.8$~eV.   In order to
assess the reliability of the computed delays, we use a reduced
close-coupling expansion by excluding the N- and (N+1)-electron CSF
involving the $4s$ correlation orbital,
which is responsible for this artifact. In this truncated model, the
energy scale is shifted downward by $0.45$~eV to match the Ar $3s$ photoemission
threshold. Another pseudo-resonance remains at $41.3$~eV, 
affecting the Wigner-like
delay at energies close the sideband 26,
preventing comparison with the last experimental value.

In Fig.~\ref{fig:3s3pdel}~(a) we present the Wigner-like delay obtained with the reduced model. 
It is computed by Eq.~(\ref{Wignerlikedelay}) for channels
leading to photoemission from the $3p$ and $3s$ orbitals.  In
Fig.~\ref{fig:3s3pdel}~(b) we show the delay difference between the two
dominant channels, $3s3p^6\epsilon p$ and $3p^5\epsilon d$,  using the
full and reduced close-coupling models. An exact determination of $\tau_{cc}$ is difficult and we, therefore, compare these curves to the original experimental data points for the Wigner-like delay of Ref.~\cite{Gueetal:12a} together with the ones estimated with the alternative $\tau_{cc}$ from Eq.~(\ref{softphotonccdelay}) and Ref.~\cite{IvaSmi:11a}.
There is good agreement between our theoretical model and the experimental points of sidebands 22 and 24.

Going beyond the present analysis requires to include the many shake-up thresholds opening at
these energies~\cite{Kiketal:96a}, as shown in Fig.~\ref{fig:levels}. As shown in experimental data \cite{Mobetal:93a} and in R-matrix
calculations~\cite{HarGre:98a}, the $3s$ photoionization 
partial and the total cross-sections are highly structured
by resonances associated to those thresholds. This suggests that the
corresponding phases may be strongly altered as well.
However, the widths of these resonances are in general below
0.1~eV so that we can expect that they do not drastically affect the measured
delays obtained from an average over the sidebands. 
Still, as we will show below, some structure should be observable provided that the IR probe field is sufficiently narrow.

\medskip

The sideband amplitude can be obtained as the convolution of the IR spectral envelope with the dipole element between the perturbed wave function and the final state~\cite{Dahetal:12b}. If, as before, we limit our discussion to Eq.~(\ref{simpledelays}), we can estimate the effect of the probe field bandwidth $\Gamma_{IR}$
by extracting the Wigner-like delay from the complex amplitudes convoluted with a gaussian function
\beq\label{eq:ampsmooth}
\tilde A_\alpha(\omega) = \sqrt{\frac{4\text{ln}2}{\pi\Gamma_{IR}^2}}\int\!\!dx\ A_\alpha(x) \text{exp}\left[-\frac{4\text{ln}2(\omega-x)^2}{\Gamma_{IR}^2}\right].
\eeq

\begin{figure}
\caption{ Relative Wigner-like delays using a model including some doubly excited resonances (see Appendix: A). (a) The probe field bandwidth is 60~meV and 0.6~eV, respectively, for the top and bottom panels. Length (continuous red) and velocity (dashed blue) gauge results are showed. The thicker, light curves correspond to the reduced model used in Fig.~\ref{fig:3s3pdel}. Experimental points are showed with the same notation as in Fig.~\ref{fig:3s3pdel}.\label{fig:delsmooth} }
\includegraphics[width=\columnwidth]{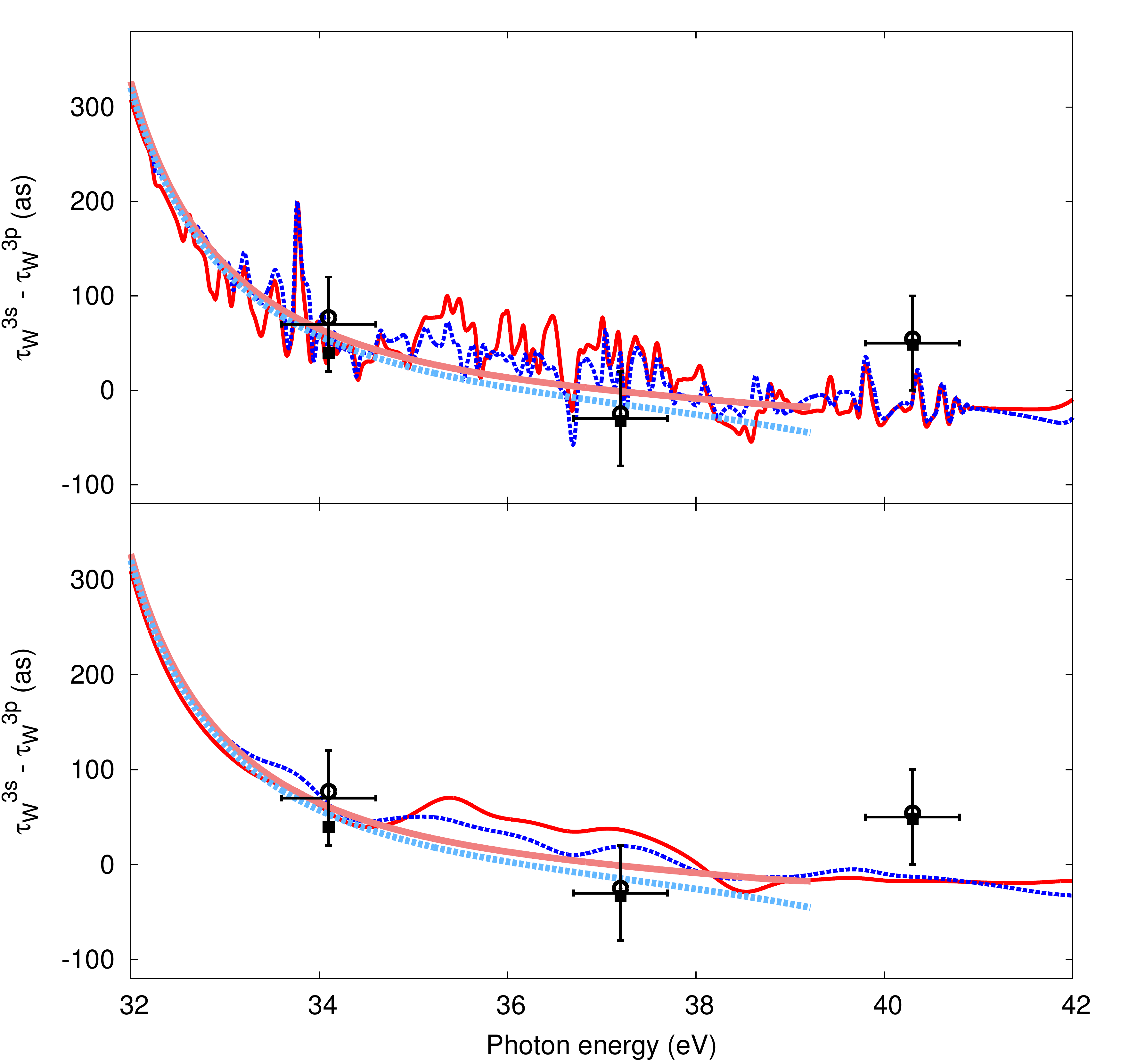} \end{figure}

In Fig.~\ref{fig:delsmooth}, we present Wigner-like delay differences obtained with a model that includes a range of physical resonances associated to the thresholds in black in Fig.~\ref{fig:levels} (details are found in Appendix~\ref{appendix}), after smoothing by the bandwidth of the probe field using Eq.~(\ref{eq:ampsmooth}). The top panel shows the results for a gaussian IR envelope with FWHM of 60~meV, corresponding to an IR field duration of 30~fs, typical of  experiments using trains of attosecond pulses. The bottom panel corresponds to a FWHM of 0.6~eV, corresponding to a single IR cycle probe pulse, typical of experiments using single attosecond pulses. We show that the presence of resonances does not change radically the baseline of the relative delays, but rather superimpose a structure that preserves the previous theory-experiment agreement. Including the physical resonances in the model also pushes pseudo-resonances higher in the spectrum so that a comparison can be made with the experimental delay extracted at the sideband 26. Here, some limited agreement can still be found in the top panel, within the theoretical and experimental uncertainty. The Cooper minimum emphasized in Ref.~\cite{Gueetal:12a}, experimentally located at about 43~eV, is also located at higher photon energies in our calculations so that it does not affect the computed delays.
We should also note that, above 39~eV, many more resonances could affect $\tau^\alpha_W$, corresponding to shake-up electrons in higher orbitals, even though they seem to be less important as far as the cross-section is concerned~\cite{Mobetal:93a}.

An important implication of these results is that a rich structure, originating from resonances, could be observed experimentally by scanning the energy of the high-order harmonics, see Fig.~\ref{fig:delsmooth}~(a). In order to observe such features,  the bandwidth of the IR field should be sufficiently narrow. Otherwise, if a too short probe pulse is used, which is often the case in attosecond streaking experiments, these sharp energy-dependent features cannot be resolved. 
In this case, the resonances may affect the measured delay smoothly, 
as is shown in Fig.~\ref{fig:delsmooth}~(b), 
where they cause an additional delay of up to several tens of attoseconds 
as compared to the model without resonances in this energy region.

Such a streaking experiment has been performed in neon to extract a relative photoionization delay of 21(5)\,as between the $2p$ and $2s$ electron extraction \cite{Schultze2010a}. No theoretical analysis has been able reproduce this result so far. Two of them include the effects of a probe field within an \emph{ab initio} many-body framework. Full-fledged simulations of the measurement have been performed by Moore et al~\cite{Mooetal:11a}, using time-dependent R-matrix theory, resulting in a relative delay of 10.2(1.3)\,as. Later, we have reproduced this result using two-photon matrix elements computed using many-body perturbation theory \cite{Dahetal:12c}. In that study, the validity of Eq.~(\ref{simpledelays}) is demonstrated. In both theoretical approaches the effects of transient double excitations are neglected.
As in the case of Argon, the Ne$^+$ level structure at the relevant energy, 
i.e. a XUV energy of about $105$~eV, is extremely rich~\cite{Kiketal:96a}. 
Here we have seen that such effects might play an important role. 
In Neon, an extra complication is that the experimental photon energy is above several double ionization thresholds, and that triply excited states may also be important, as pointed out in Ref.~\cite{Mooetal:11a}. 
This implies that it will be important  to include a complete description of at least two active electrons, and possibly excited configurations of the grand-parent system. 


\section{Conclusions and Outlook}\label{sec:conclusions}

We have presented a practical method, based on the ATomic Structure Package
(ATSP2K)~\cite{Froetal:07a}, to obtain a general, many-electron
atomic wave function in a close-coupling expansion. The structure of the localized part of the wave function is represented using
the Multi-Configuration Hartree-Fock (MCHF) method, which provides a versatile and flexible way to represent complex many-body effects with a limited number of configurations. By using B-splines to describe the radial components of the 
photoelectron in the allowed partial-wave channels, we obtain a multi-channel method that, for the description of one active electron in a box, is complete.
This paper develops this approach in combination with the exterior
complex scaling method of use, in particular, in the description of monochromatic
wave packets.
In addition to evaluating  the  total cross-section with the standard tools of complex scaling, we have extracted complex partial amplitudes that encodes the full information on the studied process. The consistency between the two approaches has been verified in detail.

The method is validated by application to the photoionization of Argon close to the $3s$ ionization threshold. In this energy region the final state is affected by strong configuration interaction.
We found good agreement of the line profiles of the $3s3p^6np$ autoionizing states with previous R-matrix calculations and experiments. Also the computed photoelectron angular distribution is in excellent agreement with the measurements.

It has earlier been  suggested that laser-assisted photoionization by attosecond pulses
gives access, in a universal way, to relative one-photon Wigner-like delays (see e.g. Refs.~\cite{Schultze2010a,klunder:2011,Gueetal:12a}). So far, this interpretation of the observations has remained uncertain since no \emph{ab initio} calculation have been able to reproduce the experimental data, even though some encouraging results were obtained in the case of Argon~\cite{Gueetal:12a,Dahetal:12c}.
In this paper, we present a study that contributes to validate these claims, in the context of a weak probe field,
by including all relevant correlation effects, and by showing that, in so doing, satisfactory agreement is obtained with the observed atomic delay shifted by the atom-independent continuum-continuum delay.
We also show that the presence of many resonances decaying into  Ar$^+\left( 3p^5 \right)$ and Ar$^+\left( 3s3p^6 \right)$  in the studied energy region affects the result of the measurements in ways that are sensitive to experimental parameters. In a future study we will extend the treatment to include consistent \emph{ab initio} calculations of the single photon and laser-assisted photoelectron emission delays, in a similar way as done is Ref.~\cite{Dahetal:12c}.

However, it has been shown for the helium atom that dynamical screening effects caused by the polarization of the parent ion by the IR field can affect the delay of electron emission~\cite{Pazetal:12a}. The rich structure of the Ar$^+$ spectrum at the relevant energies~\cite{Kiketal:96a} suggests that such an effect could be significant in the present case. A straightforward possibility to include those effects in the calculation is to move to a time-dependent approach, including explicitly the shape of the IR pulse in order to account for the full
many-body dynamics of the process, both on the relatively long time-scale of  autoionization, and the 
sub-femtosecond scale set by the XUV-pulse.

For further developments we note that
the compactness of the MCHF and B-spline close-coupling approach presented here makes it well suited for 
time-dependent many-body calculations on systems in which exact calculations are untractable.  Only a limited number of such calculations exist today, see for example~\cite{Haretal:07a,Guan2008,Lysetal:09a,Hutetal:11a,Mooetal:11a}, all of them performed within the Time-Dependent R-matrix framework~\cite{Burke1997,Lysetal:09a,Guan2007}. 
As a next step we will combine the present approach with a solution of the time-dependent Schr{\"o}dinger equation
along the lines of Ref.~\cite{argenti:prl:10,Argetal:12a}, where the calculation is made without complex scaling and 
the final physical result are  extracted by projection of the finite-time wave packet on K-matrix scattering states. Absorbing boundaries are then used in order to avoid reflections on the walls of the computational box
in the case of long simulations. We foresee here that the common description of the structure part in the 
time-independent and time-dependent calculation will be of significant value for the 
tuning  of the  approximations  regarding many-body effects that has to be introduced in  the time-dependent approach.

\section*{Acknowledgements}
\noindent
We thank Prof. Nora Berrah and Dr. Burkhard Langer for providing us with
experimental data.  Financial support from the Swedish Research Council
(VR) and the Wenner-Gren Foundation is gratefully acknowledged.

\appendix
\section{}\label{appendix}

Here, we give the details of the calculations used to produce Fig.~\ref{fig:delsmooth}.
The model is aimed at a qualitative description of the spectral features in the range of the experiments performed in Refs.~\cite{klunder:2011,Gueetal:12a}.

According to Ref.~\cite{HarGre:98a}, the most prominent features in the $3s$ photoionization partial cross-section and total cross-section, come from the resonances of the type $3p^4nl[LS\pi]n'l'$ with $nl=3d,\, 4s,$ and $4p$, $n'\leq 7$, and for the parent term $LS\pi=\ ^2S^e, \, ^2P^o, \, ^2P^e$ and $^2D^e$. The $^2P^e$ parent states are mostly responsible for resonances at 31~eV and below so we do not consider them.
Using the same $1s,2s,2p,3s$ and $3p$ orbitals as before, we perform Hartree-Fock level average calculations on the $3p^4nl$ states optimizing the $3d,4s$ and $4p$ orbitals.
Extending the active set of spectroscopic orbitals by a $\overline{4d}$ correlation orbital, we perform a MCHF calculation on the lowest state of the $^2S^e$ symmetry of Ar$^+$ including all single and double (SD) excitations of the multi-reference set (MR)
$\{3s3p^6 \cup 3p^44s \cup 3p^43d \cup 3p^4\overline{4d}\}$. Configuration interaction calculations are then performed on the $^2P^o$ and $^2D^e$ states following the same MR-SD scheme with the references $\{3p^44s \cup 3p^43d \cup 3p^4\overline{4d}\}$ and $\{3p^5 \cup 3p^44p\}$, respectively. Then, the CSF with a mixing coefficient over $0.07$ are selected to build the target states. The lowest levels of Ar$^+$ obtained in this model are compared to experiment in Table~\ref{tab:Ar+spec}. Most values agree with the experiment within 1~eV.

\begin{table}\caption{Comparison of the lowest computed levels of Ar$^+$ to R-matrix calculations~\cite{HarGre:98a} and experimental data~\cite{Min:63a}.\label{tab:Ar+spec}}
\begin{ruledtabular}
\begin{tabular}{rD{.}{.}{5}D{.}{.}{5}D{.}{.}{5}}
\multicolumn{1}{c}{State} & \multicolumn{1}{c}{this work} & \multicolumn{1}{c}{Ref.~\cite{HarGre:98a}} & \multicolumn{1}{c}{Exp.~\cite{Min:63a} }\\
\hline
$3p^5        \ ^2P^o $& 16.481 & 16.303 & 15.819\\
$3s3p^6      \ ^2S^e $& 29.24  & 29.24  & 29.24 \\
$3p^4(^1D)4s \ ^2D^e $& 34.699 & 34.999 & 34.203\\
$3p^4(^3P)3d \ ^2D^e $& 35.418 & 35.427 & 34.462\\
$3p^4(^3P)4p \ ^2P^o $& 36.044 & 36.514 & 35.605\\
$3p^4(^1S)4s \ ^2S^e $& 37.394 & 37.207 & 36.504\\
$3p^4(^1D)4p \ ^2P^o $& 37.654 & 38.134 & 37.137\\
$3p^4(^1D)3d \ ^2D^e $& 38.647 & 38.508 & 37.148\\
$3p^4(^1S)3d \ ^2D^e $& 39.498 & 39.296 & 38.043\\
$3p^4(^1D)3d \ ^2S^e $& 39.595 & 39.876 & 38.585\\
\end{tabular}
\end{ruledtabular}
\end{table}

The localized states $\chi_\xi$ are chosen such that the close-coupling expansion Eq.~(\ref{ciansatz}) still retains the completeness of the B-spline set.
Using this model, the photon energy scale is shifted by $0.15$~eV upward in order to match the experimental $3s$ photoionization.
We choose a knot sequence as before for the inner region up to 1/Z~$a_0$, but with a linear step of 1~$a_0$ up to $51$~$a_0$ and then an exponentially increasing step up to $150$~$a_0$ for accommodating orbitals with $n\leq7$. The complex scaling starts at $40$~$a_0$ with a complex scaling angle of $0.11$~rad. As in Section~\ref{sec:res}, there is only a negligible difference between the sum of the partial cross-sections and the total cross-section computed with Eq.~(\ref{xsecCS}).
In Table~\ref{tab:resapen}, we provide the results obtained for the $3s3p^6np$ resonances line shapes. A good consistency is found with the values given in Table~\ref{restab}, except for the $q$ parameters. Furthermore, the partial cross-sections obtained within this model exhibit strong features in qualitative agreement with R-matrix calculations~\cite{HarGre:98a}.

\begin{table}\caption{Computed Fano profile parameters for the $3s3p^6np$ resonances. Positions are shifted upward by 0.15~eV.\label{tab:resapen}}
\begin{ruledtabular}
\begin{tabular}{cD{.}{.}{3}D{.}{.}{1}D{.}{.}{3}D{.}{.}{3}}
$n$ &
	\multicolumn{1}{c}{$\Delta E$} & \multicolumn{1}{c}{$\Gamma$} &
	\multicolumn{1}{c}{$q$ (L)} & 	\multicolumn{1}{c}{$q$ (V)} \\
& \multicolumn{1}{c}{eV} & \multicolumn{1}{c}{meV}\\
\hline
4 & 26.524 & 81.0 & -0.497 & -0.646 \\
5 & 28.005 & 25.0 & -0.435 & -0.489 \\
6 & 28.516 & 11.7 & -0.409 & -0.443 \\
7 & 28.764 &  6.4 & -0.396 & -0.423 \\
8 & 28.903 &  3.9 & -0.389 & -0.412 
\end{tabular}
\end{ruledtabular}
\end{table}

\bibliography{library,cc} \end{document}